\documentclass[preprintnumbers,amsmath,amssymb,floatfix,11pt,aps,onecolumn,
groupedaddress,nofootinbib]{revtex4-1}
\usepackage{latexsym}
\usepackage{epsfig}
\usepackage{epstopdf}
\usepackage{amssymb}
\usepackage{mathtools}
\usepackage{amsmath,hyperref}

\begin{document}
\title{Rotating normal and phantom Einstein--Maxwell--dilaton black holes: Geodesic analysis}

\author{Mustapha Azreg-A\"{i}nou}
\email{azreg@baskent.edu.tr}\affiliation{Ba\c{s}kent University, Engineering Faculty, Ba\u{g}l\i ca Campus, Ankara, Turkey}
\author{Sumarna Haroon}
\email{sumarna.haroon@sns.nust.edu.pk}\affiliation{Department of Mathematics, School of Natural
	Sciences (SNS), National University of Sciences and Technology
	(NUST), H-12, Islamabad, Pakistan}\affiliation{Department of Physics and Astronomy, University of Waterloo, Waterloo, Ontario, Canada}
\author{Mubasher Jamil}
\email{mjamil@sns.nust.edu.pk}\affiliation{Department of Mathematics, School of Natural
	Sciences (SNS), National University of Sciences and Technology
	(NUST), H-12, Islamabad, Pakistan}
\author{Muhammad Rizwan}
\email{m.rizwan@sns.nust.edu.pk}\affiliation{Department of Mathematics, School of Natural
	Sciences (SNS), National University of Sciences and Technology
	(NUST), H-12, Islamabad, Pakistan}\affiliation{Department of Computer Science, Faculty of Engineering \& Computer Sciences, National University of Modern Languages, H-9, Islamabad, Pakistan}

\begin{abstract}
Depending on five parameters, rotating counterparts of Einstein--Maxwell--dilaton black holes are derived. We discuss their physical and geometric properties and investigate their null and time-like geodesics including circular orbits. The Lense--Thirring effect is considered.
\end{abstract}

\maketitle
\section{Introduction}
Numerous astrophysical observational programs~\cite{JL,H,N} point to the convincing conclusion that our Universe is
expanding with acceleration. The leading candidate behind this novel expansion is named as {\textquoteleft dark energy\textquoteright}. The two main characteristic
behaviors of dark energy are: (i) it has negative pressure, (ii) it violates one of the four energy conditions. The attempts to study dark energy had led to the formulation of various effective field theories. One of such theories is gravity coupled to both linear electromagnetism and a phantom dilatonic field the kinetic energy term of which has the {\textquoteleft wrong\textquoteright} sign. This is the so-called Einstein--Maxwell--anti-dilaton theory (EM$\bar{\text{D}}$).

Spherically symmetric dilaton black hole solutions were discussed in~\cite{BS,G82,GM,GHS,HPro,GS95}. Gibbons and Rasheed derived a family of black hole solutions to the enlarged class of Einstein--Maxwell--dilaton theories including EMD, E$\bar{\text{M}}$D, EM$\bar{\text{D}}$, and E$\bar{\text{M}}\bar{\text{D}}$~\cite{GR}. For short we will call it EMD theory. Their results, giving rise to new structure of black holes, called phantom black holes, were further studied in Refs.~\cite{Gao,G,P}. Regular phantom black holes have been discussed in Ref.~\cite{regular}, in the absence of a cosmological constant, and in Ref.~\cite{cosmo}, in the presence of a cosmological constant.

The purpose of the present work is to formulate a rotating black hole solution of a static EMD black hole. Many methods have been developed in theory to compute rotating solutions from static ones and the most widely known method is the Newman--Janis algorithm and its generalizations. After the mathematical formulation of Kerr black hole \cite{RP}, Newman and Janis \cite{NJ1} proposed a technique based on certain complex transformations through which the Kerr metric can be easily computed. In the same year, the Kerr-Newman black hole solution was determined using this technique \cite{NJ2}. Afterwards the Newman--Janis algorithm received a lot of fame and became a powerful tool for seeking rotating black hole solutions from their respective static counterparts. Though this formalism was analyzed very critically and was referred to as a {\textquoteleft trick\textquoteright} by some workers. Soon later it was shown how this algorithm is mathematically correct~\cite{Til,drake}. With time the Newman--Janis algorithm evolved and some ambiguities associated with this method were removed~\cite{A}. Recently the algorithm has been used in its most general form in Ref.~\cite{2017} providing the reader with prescriptions for obtaining the other fields associated with the rotating metric. To compute the rotating metric for normal or phantom static EMD black hole we have followed the Newman--Janis formalism along with the modification suggested in Ref.~\cite{A}. 

The structures of the horizons and ergosphere are discussed and it is found that the increase in the value of the rotational parameter $a$ affects the shape of the black hole. The singularities of the metric for both the $\cosh$ and $\sinh$ solutions are discussed. We have also observed that the area and entropy of the rotating phantom black hole depends on the coupling parameters of the (enlarged) EMD theory.

The geometry of the gravitational source can be well exhibited through studying its essential features, one such feature is geodesic motion around the gravitating body. Immense work has been done on studying geodesic motion of particles, both massive and photons, around black holes. A comprehensive study of static (Schwarzschild), charged (Reissner--Nordstr\"{o}m) and rotating (Kerr) black holes have been provided by Chandrasekhar~\cite{Chandra}, where he has utilized Lagrangian and Hamilton--Jacobi approaches to analyzed both time-like and null geodesics. Following the Lagrangian approach we have also investigated null and time-like geodesics for our rotating black hole metric.

This paper is section-wise organized. Section~\ref{secrev} is a brief review of normal and phantom EMD black holes, in which we recall the static spherically symmetric EMD black hole solution derived in Ref.~\cite{G}. In Sec.~\ref{secrot}, we present a rotating version of static spherically symmetric EMD black holes and in Sec.~\ref{secpgp} we discuss their physical and geometric properties. The null and time-like geodesic motion in equatorial plane is considered in Sec.~\ref{secgeod} along with the study of effective potential. The Lense--Thirring effect is discussed in Sec.~\ref{secLT}. The results are summarized in Sec.~\ref{secdis}.

\section{Normal and phantom black holes in EMD theory\label{secrev}}
The Einstein--Maxwell--dilaton action is given by~\cite{G}
\begin{equation}\label{emd}
S=\int{d^4x}\sqrt{-g}\left[\mathcal{R}-2\eta_1g^{\mu\nu}\nabla_\mu\varphi\nabla_\nu\varphi+\eta_2{\rm e}^{2\lambda\varphi}F^{\mu\nu}F_{\mu\nu}\right],
\end{equation}
where $\mathcal{R}$ represents the scalar curvature, $\varphi$ is the dilaton field, $F_{\mu\nu}$ is the electromagnetic field tensor, and $\lambda$ stands for a coupling parameter. The nature of the fields depend on the constant parameters $\eta_1$ and $\eta_2$: Normal EMD corresponds to $\eta_2=\eta_1=+1$, while phantom couplings of the dilaton field or/and Maxwell field are obtained for $\eta_1=-1$ or/and $\eta_2=-1$ yielding the theories EM$\bar{\text{D}}$ ($\eta_2=+1,\,\eta_1=-1$), E$\bar{\text{M}}$D ($\eta_2=-1,\,\eta_1=+1$), and E$\bar{\text{M}}\bar{\text{D}}$ ($\eta_2=-1,\,\eta_1=-1$).

Normal and phantom static black hole solutions in EMD theory were derived in Ref.~\cite{G} and are given by
\begin{eqnarray}\label{1}
  ds^2 &=& f(r)dt^2-\frac{dr^2}{g(r)}-h(r)d\Omega^2,
\end{eqnarray}
where
\begin{eqnarray}
\label{fg} f(r)&=&g(r)=\Big(1-\frac{r_1}{r}\Big)\Big(1-\frac{r_2}{r}\Big)^\gamma, \\
\label{hr}  h(r)&=&r^2 \Big(1-\frac{r_2}{r}\Big)^{1-\gamma},
\end{eqnarray}
and

\begin{equation}\label{gdom}
\lambda_{\pm}=1\pm \eta_1\lambda^2,\qquad \gamma=\frac{\lambda_-}{\lambda_+}\in 
 \begin{cases}
  (-\infty,-1)\cup[1,\infty)\quad&\text{if}\quad \eta_1 = -1 \\
   (-1,1]\quad&\text{if}\quad\eta_1 =+1
  \end{cases}.
\end{equation}
The two horizons of this black hole are given by
\begin{align}
\label{hor}&r_1=M+\sqrt{M^{2}-\frac{2\eta_{2}\gamma{q}^2}{1+\gamma}}, & & r_2=\frac{2\eta_2q^2}{(1+\gamma)r_1}=\frac{1}{\gamma}\left(M-\sqrt{M^{2}-\frac{2\eta_{2}\gamma{q}^2}{1+\gamma}}\right), & (\gamma\neq 0),\\
\label{hor2}&r_1=2M, & & r_2=\frac{\eta_2q^2}{M}, & (\gamma = 0),
\end{align}
with the conditions
\begin{align}
\label{cd1}& r_2<0<{r_1}& &{\text{for}}\quad \frac{\eta_2}{1+\gamma}<0,\\
\label{cd2}&0<{r_2}\leq{r_1}& &{\text{for}}\quad \frac{\eta_2}{1+\gamma}>0\;\text{ and }\;M^{2}\geq \frac{2\eta_{2}\gamma{q}^2}{1+\gamma},\\
&0=r_2<r_1=2M& &{\text{for}}\quad \eta_2q^2=0.
\end{align}
Here $M$ is the mass and $q$ is the electric charge of the black hole. The cosh and sinh solutions aree defined by
\begin{eqnarray}
\label{cd3}  \eta_2(1+\eta_1\lambda^2)=\frac{2\eta_2}{1+\gamma}<0\quad &\text{for cosh solution}&,\\
\label{cd4}  \eta_2(1+\eta_1\lambda^2)=\frac{2\eta_2}{1+\gamma}>0\quad &\text{for sinh solution}&.
\end{eqnarray}
The cosh solution has no extremal black hole. The sinh solution has an extremal black hole if subjected to the constraints: $2M^2=\eta_2(1+\gamma)q^2$ and $\eta_2=+1$. When this is the case, the extremal black hole has a double horizon at $r_{\text{ext}}=2M/(1+\gamma)$.

The associated electromagnetic $F$ and scalar $\Phi$ fields are given by
\begin{equation}\label{ems}
F=-\frac{q}{r^2}\,d r\wedge d t\, ,\;\;
e^{-2\lambda\Phi}=\Big(1-\frac{r_2}{r}\Big)^{1-\gamma}.
\end{equation}

The above solution is a family of static, spherically symmetric and asymptotically flat black holes. Our next aim is to formulate a rotating counterpart of the metric (\ref{1}).

\section{Derivation of rotating normal and phantom black holes without complexification\label{secrot}}
The first step of the Newman--Janis algorithm is to transform from Boyer--Lindqiust coordinates $(t,r,\theta,\phi)$ to Eddington--Finkelstein coordinates $(u,r,\theta,\phi)$. On applying the coordinate transformation $dt=du+dr/\sqrt{f(r)g(r)}$ to Eq. (\ref{1}) we obtain
\begin{eqnarray}\label{b}
ds^{2}=f(r)du^{2}+2\sqrt{\frac{f(r)}{g(r)}}dudr-h(r)d\theta^{2}-h(r)\sin^{2}\theta{d\phi}^2.
\end{eqnarray}
This metric can be represented in terms of null tetrads~\cite{A}
\begin{eqnarray}\label{d}
g^{ab}=l^{a}n^{b}+l^{b}n^{a}-m^{a}\bar{m}^{b}-m^{b}\bar{m}^{a},
\end{eqnarray}
where
\begin{eqnarray}
l^{a}&=&\delta^{a}_{r},\nonumber\\
\label{e0} n^a&=&\sqrt\frac{g(r)}{f(r)}\delta^{a}_{u}-\frac{g(r)}{2}\delta^{a}_{r},\\
m^a&=&\frac{1}{\sqrt{2h(r)}}(\delta^{a}_{\theta}+\frac{\dot{\iota}}{\sin\theta}\delta^{a}_{\phi}).\nonumber
\end{eqnarray}
These null tetrads satisfy the following conditions
\begin{eqnarray}\nonumber
l^{a}l_{a}=n^{a}n_{a}=m^{a}m_{a}=\bar{m}^{a}\bar{m}_{a}=0,\\
l^{a}m_{a}=l^{a}\bar{m}_{a}=n^{a}m_{a}=n^{a}\bar{m}_{a}=0,\\\label{f}\nonumber
l^{a}n_{a}=-m^{a}\bar{m}_{a}=1.
\end{eqnarray}
Now we apply the second step of the Newman--Janis algorithm which consists in performing the complex coordinate transformation in the $ur$-plane
\begin{eqnarray}\nonumber
u'\rightarrow{u-\dot{\iota}{a}\cos\theta},\\\label{g}
r'\rightarrow{r+\dot{\iota}{a}\cos\theta},
\end{eqnarray}
where $a$ is the rotational parameter.

The third step of the Newman--Janis algorithm consists in complexifying the radial coordinate $r$: There are as many ways to complexify $r$ as one wants and this is very ambiguous as shown in Ref~\cite{A}. One of us has resorted to a new procedure~\cite{A,A1} by which one drops the complexification step of the Newman--Janis algorithm. The procedure has known applications in a series of papers~\cite{s1,s2,s3,s4,s5,s6,s7,s8,s9,s10,s11}.

In the new procedure we admit that $\delta^{\mu}_{\nu}$, in Eq.~\eqref{e0}, transform as a vector under the transformation~\eqref{g} and that the functions $f(r)$, $g(r)$ and $h(r)$ transform to $F=F(r,a,\theta)$, $G=G(r,a,\theta)$ and $H=H(r,a,\theta)$ respectively. Thus our new null tetrads are
\begin{eqnarray}
l^{a}&=&\delta^{a}_{r},\nonumber\\
\label{e} n^a&=&\sqrt\frac{G}{F}\delta^{a}_{u}-\frac{G}{2}\delta^{a}_{r},\\\
m^{a}&=&\frac{1}{\sqrt{2H}}[(\delta^{a}_{u}-\delta^{a}_{r})\dot{\iota}{a}\sin\theta+\delta^{a}_{\theta}+\frac{\dot{\iota}}{\sin\theta}\delta^{a}_{\phi}].\nonumber
\end{eqnarray}
Using these null tetrads, the contravariant components of the rotating metric are given by Eq.~(9) of Ref.~\cite{A}, which we rewrite using the notation of the present work (the correspondence between the two notations Ref.~\cite{A} $\to$ present work is: $G(r)\to f(r)$, $F(r)\to g(r)$, and $H(r)\to h(r)$ for the nonrotating solution and $A(r,\theta)\to F(r,\theta)$, $B(r,\theta)\to G(r,\theta)$, and $\Psi(r,\theta)\to H(r,\theta)$ for the rotating solution)
\begin{align*}
&g^{uu}=\frac{-a^{2}\sin^{2}\theta}{H},\qquad g^{u\phi}=\frac{-a}{H},\qquad
g^{ur}=\sqrt{\frac{G}{F}}+\frac{a^{2}\sin^{2}\theta}{H},\\
&g^{rr}=-G-\frac{a^{2}\sin^{2}\theta}{H},\qquad
g^{r\phi}=\frac{a}{H},\qquad
g^{\theta\theta}=\frac{-1}{H},\qquad
g^{\phi\phi}=-\frac{1}{H\sin^2\theta}.
\end{align*}
So our new metric in Eddington--Finkelstein coordinates is [Eq.~(10) of Ref.~\cite{A}]
\begin{eqnarray}
ds^2&=&F d u^2+2 \frac{\sqrt{F}}{\sqrt{G}} d u d r+2 a\sin ^2\theta  \Big(\frac{\sqrt{F}}{\sqrt{G}}-F\Big) d u d \phi \nonumber\\
&-& 2 a\sin ^2\theta  \frac{\sqrt{F}}{\sqrt{G}} d r d \phi-H  d \theta^2 - \sin ^2\theta  \Big[H +a^2\sin ^2\theta  \Big(2 \frac{\sqrt{F}}{\sqrt{G}}-F\Big)\Big] d \phi^2.
\end{eqnarray}
The final but crucial step is to bring this form of the metric to Boyer--Lindquist coordinates by a global coordinate transformation of the form
\begin{eqnarray}
du&=&dt+\lambda(r)dr,\\\nonumber
d\phi&=&d\phi'+\chi(r)dr,
\end{eqnarray}
where~\cite{A,A1}
\begin{equation}\label{k}
{\lambda}=-\frac{a^2+k(r)}{a^2+g(r)h(r)},\qquad~ {\chi}=-\frac{a}{g(r)h(r)+a^2},\qquad~ k(r)={\sqrt\frac{g(r)}{f(r)}}h(r).
\end{equation}
Since the function $F$, $G$, and $H$ are still unknown, one can fix some of them to get rid of the cross term $dtdr$ in the metric. This is generally not possible in the usual Newman--Janis algorithm since these functions are fixed once the complexification of $r$ is performed and there remains no free parameters or functions to act on to achieve the transformation to Boyer--Lindquist coordinates.

Now, if we choose~\cite{A,A1}
\begin{equation}\label{FG}
F=\frac{\left(g(r)h(r)+a^2\cos^2\theta\right)H}{\left(k(r)+a^2\cos^2\theta\right)^2},\qquad G=\frac{\left(g(r)h(r)+a^2\cos^2\theta\right)}{H},
\end{equation}
the rotating black hole solution in Boyer--Lindquist coordinates turns out to be in its generic form of the Kerr-like metric [Eq.~(16) of Ref.~\cite{A}]
\begin{align}
&d s^2 =\frac{H}{k+a^2 \cos^2\theta}\Big[\Big(1-\frac{\sigma}{k+a^2 \cos^2\theta}\Big)d t^2-\frac{k+a^2 \cos^2\theta}{\Delta}\,d r^2\nonumber\\
\label{B}&+\frac{2a\sigma \sin ^2\theta}{k+a^2 \cos^2\theta}\,d td \phi-(k+a^2 \cos^2\theta)d \theta^2-\frac{[(k+a^2)^2-a^2\Delta\sin^2\theta]\sin ^2\theta}{k+a^2 \cos^2\theta}\,d \phi^2\Big],
\end{align}
where
\begin{equation}\label{D}
\sigma(r)\equiv k-gh,\qquad \Delta(r)\equiv gh+a^2,
\end{equation}
and to simplify the notation we have dropped the prime from $\phi$.

Now, by a straightforward application we generate the rotating counterpart of the EMD static metric~\eqref{1} where $f(r)=g(r)$. This implies $k(r)=h(r)$~\eqref{k}. The function $H(r,\theta,a)$ is still arbitrary in~\eqref{B} and this can be chosen so that the cross term of the Einstein tensor $G_{r\theta}$, for a physically acceptable rotating solution, identically vanishes: $G_{r\theta}\equiv 0$. The latter constraint yields [Eq.~(19) of Ref.~\cite{A}]
\begin{equation}\label{cnstrt}
\left(h+a^2 y^2\right)^2 \left(3H_{,r}H_{,y^2} -2H H_{,ry^2}\right) =3a^2h_{,r}\,H^2,
\end{equation}
where $y\equiv \cos\theta$. It is easy to see that the particular expression of $H$,
\begin{equation}\label{H}
H=h(r)+a^2y^2=h(r)+a^2\cos^2\theta,
\end{equation}
is a solution to the partial differential equation~\eqref{cnstrt}. This particular expression of $H$ yields $F=G$~\eqref{FG} and the rotating version of the phantom static black hole acquires the form [Eq.~(24) of Ref.~\cite{A}]
\begin{equation}\label{metric}
ds^2=\left(1-\frac{\sigma}{H}\right)dt^2+2a\frac{\sigma}{H}\sin^2\theta dtd\phi-\frac{H}{\Delta}dr^2-H{d\theta}^2-\sin^2\theta\left[h+a^2+a^2\frac{\sigma}{H}\sin^2\theta\right]d\phi^2,
\end{equation}
where $H$ is given by~\eqref{H}, $\sigma=h(1-f)$, and $\Delta=fh+a^2$~\eqref{D}. These latter expressions are expressed in terms of the roots $r_1$ and $r_2$~(\ref{hor}, \ref{hor2}) and $h(r)$~\eqref{hr} as
\begin{align}
\label{Ds1}&\Delta(r)=fh+a^2 =r^2-(r_1+r_2)r+r_1r_2+a^2,\\
\label{Ds2}&\sigma(r)=h(1-f) =h(r)+(r_1+r_2)r-r_1r_2.
\end{align}

The metric~\eqref{metric} reduces to the Kerr solution for $\gamma=1$ and $q=0$  and to the normal and phantom Kerr--Newman metric for $\gamma=1$, $q\neq{0}$, and $\eta_2=+1$ or $\eta_2=-1$, respectively. Despite the fact that metric~(\ref{metric}) describes rotating normal and phantom EMD black holes, for future references, we refer to it, for short, as rotating phantom black hole (RPBH).

In order to determine the electromagnetic and scalar fields of the rotating solution one has to solve the field equations (Eqs.~(2.2, 2.3, 2.4) of Ref.~\cite{G}) using the metric~\eqref{metric}, \eqref{Ds1} and~\eqref{Ds2} as an ansatz. For that end one may use decomposition methods~\cite{dec1,dec2,dec3}, a task postponed to a subsequent paper. Using continuity arguments one may claim that solutions for the electromagnetic and scalar fields exist anyway even in non-closed forms. In fact, since for $\gamma=1$, the metric~\eqref{metric}, \eqref{Ds1} and~\eqref{Ds2} provides three exact solutions to the field equations (Kerr and normal and phantom Kerr--Newman black holes), solutions for $\gamma$ in the vicinity of 1 should exist by continuity.

Those rotating solutions derived in~\cite{s1,s2,s3,s4,s5,s6,s7,s10,s11} on applying our procedure and some other solutions known in the literature~\cite{gen1} could have been obtained by mere substitution in the generic rotating solution derived in Refs.~\cite{A,A1}, which is Eq.~\eqref{B} of this manuscript. Some authors were fair in citing our \emph{original work consisting in dropping the complexification of the radial coordinate $r$}~\cite{A,A1}. Some other authors have failed to do that~\cite{s3,s10,s11}; they have borrowed many steps and equations from our \emph{original} idea of generating rotating solutions~\cite{A,A1} without, however, citing the sources. Said otherwise, using different notations they have just repeated {\textquotedblleft previously obtained results without giving proper references\textquotedblright}~\cite{comment2} and with no inferred novelty. Another important point to note is that most workers, if not all, avoid elaborating on the choice of the function $H=h+a^2\cos^2\theta$~\eqref{H}. We have shown here, as we did in Refs.~\cite{A,A1}, that such a choice is always possible if $f=g$ and yields a physically acceptable solution where $G_{r\theta}\equiv 0$ with $G_{\mu\nu}$ being the Einstein tensor.

In the following section we discuss some significant properties of the RPBH.

\section{Physical and geometrical properites of the rotating normal and phantom EMD black hole\label{secpgp}}

\paragraph{\textbf{Horizons.}}
The horizon of a black hole is defined by the requirement $g^{rr}=0)$. As is well known, the Kerr--Newman geometry has two horizons if $M^2>a^2+q^2$, a naked singularity if $M^2<a^2+q^2$, or an extremal black hole with two merging horizons if $M^2=a^2+q^2$. For the RPBH case the radii of the horizons depend not only on the mass $M$, angular momentum $a$, and charge $q$ of the black hole; rather they also depend on the coupling parameters $\eta_2$ and $\gamma$~\eqref{gdom}. From the metric (\ref{metric}) we have
\begin{eqnarray}
	g_{rr}=-\frac{H}{\Delta}.
\end{eqnarray}
The above term is singular when $\Delta=0$, this yields
\begin{equation}\nonumber
	r^2-\left(r_1+r_2\right)r+r_1r_2+a^2=0.
\end{equation}
The solutions of this equation gives the radii $r_+>r_-$ of the horizons by
\begin{eqnarray}\label{horizon}
	r_\pm&=&\frac{(r_1+r_2)\pm\sqrt{(r_1-r_2)^2-4a^2}}{2},
\end{eqnarray}
which are independent of $\theta$. This property is not valid only for~\eqref{metric}; rather, it applies to all rotating metrics~\eqref{B} for all $f$, $g$, $h$, and $H$. Here $r_1$ and $r_2$ depend on $M$, $q^2$, $\eta_2$, and $\gamma$~(\ref{hor}, \ref{hor2}). 

For the case of the normal ($\eta_2=+1$) or phantom ($\eta_2=-1$) Kerr--Newman black hole ($\gamma=1$), Eq.~\ref{horizon} reduces to the well-known expression
\begin{eqnarray}\label{horizon1}
	r_{\small{\text{KN}}^{\pm}}=M\pm{\sqrt{M^2-\eta_2q^2-a^2}}.
\end{eqnarray}
\begin{figure}
    \includegraphics[width=0.49\linewidth]{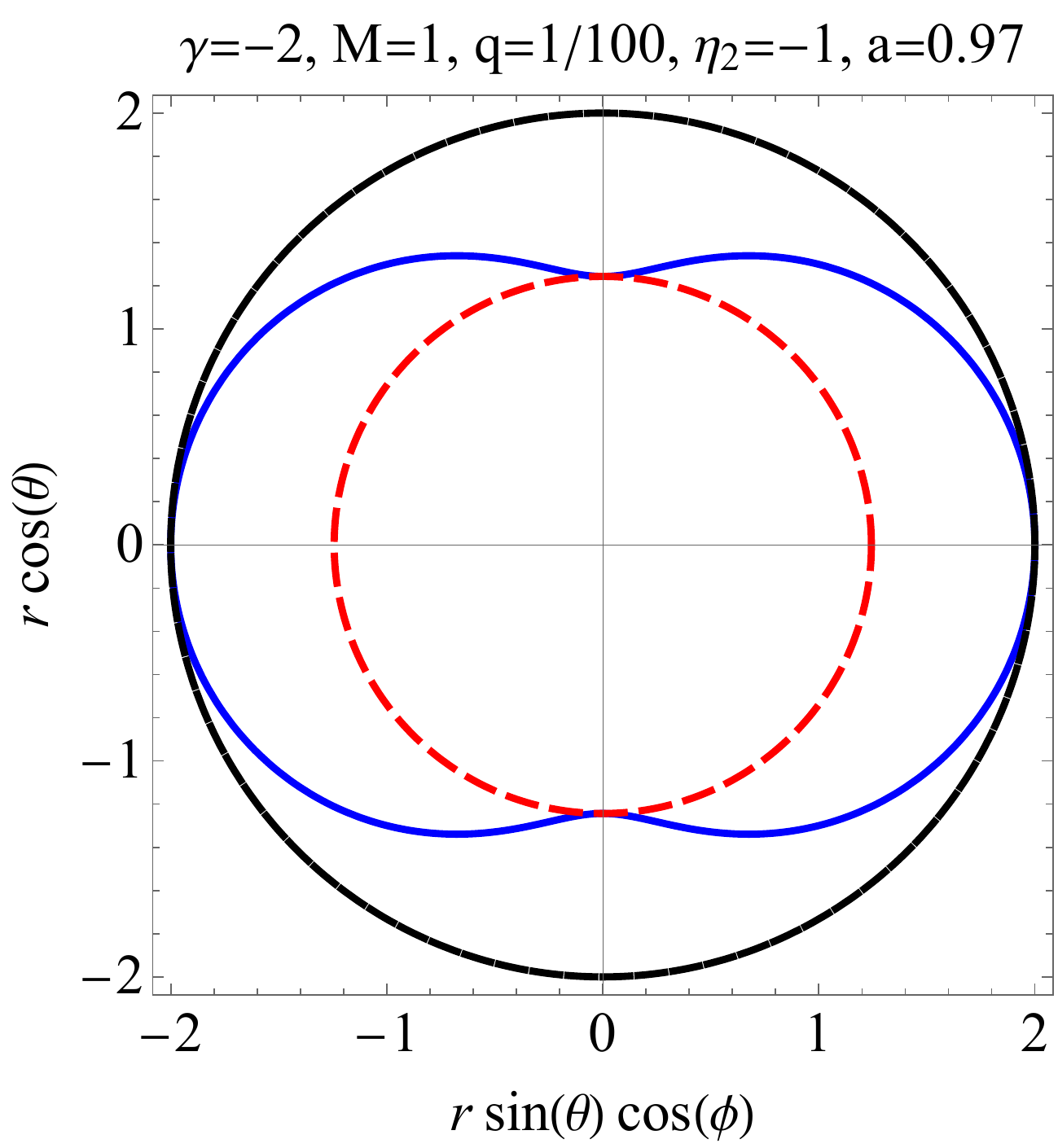}
    \includegraphics[width=0.49\linewidth]{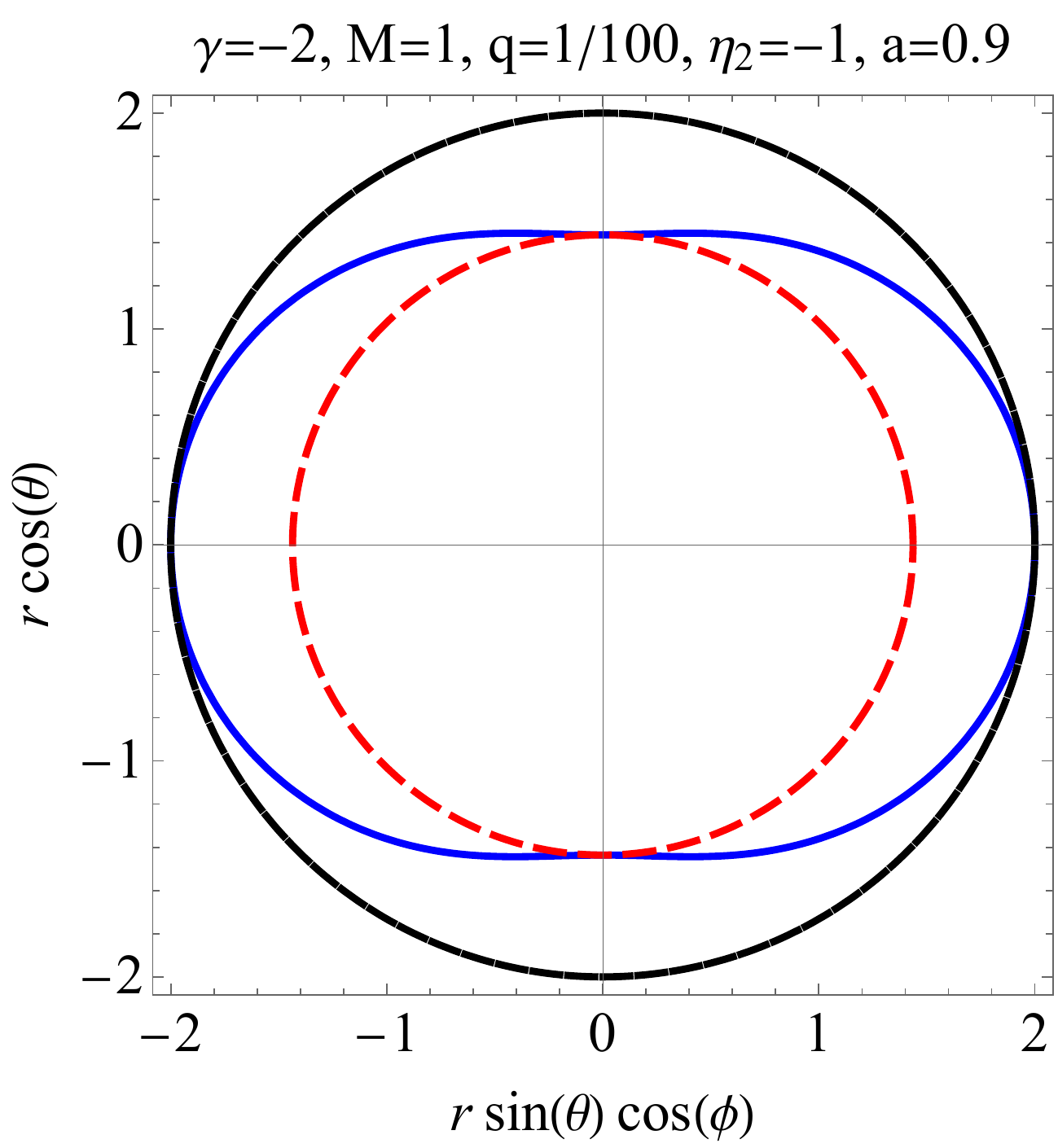}\\
    \includegraphics[width=0.49\linewidth]{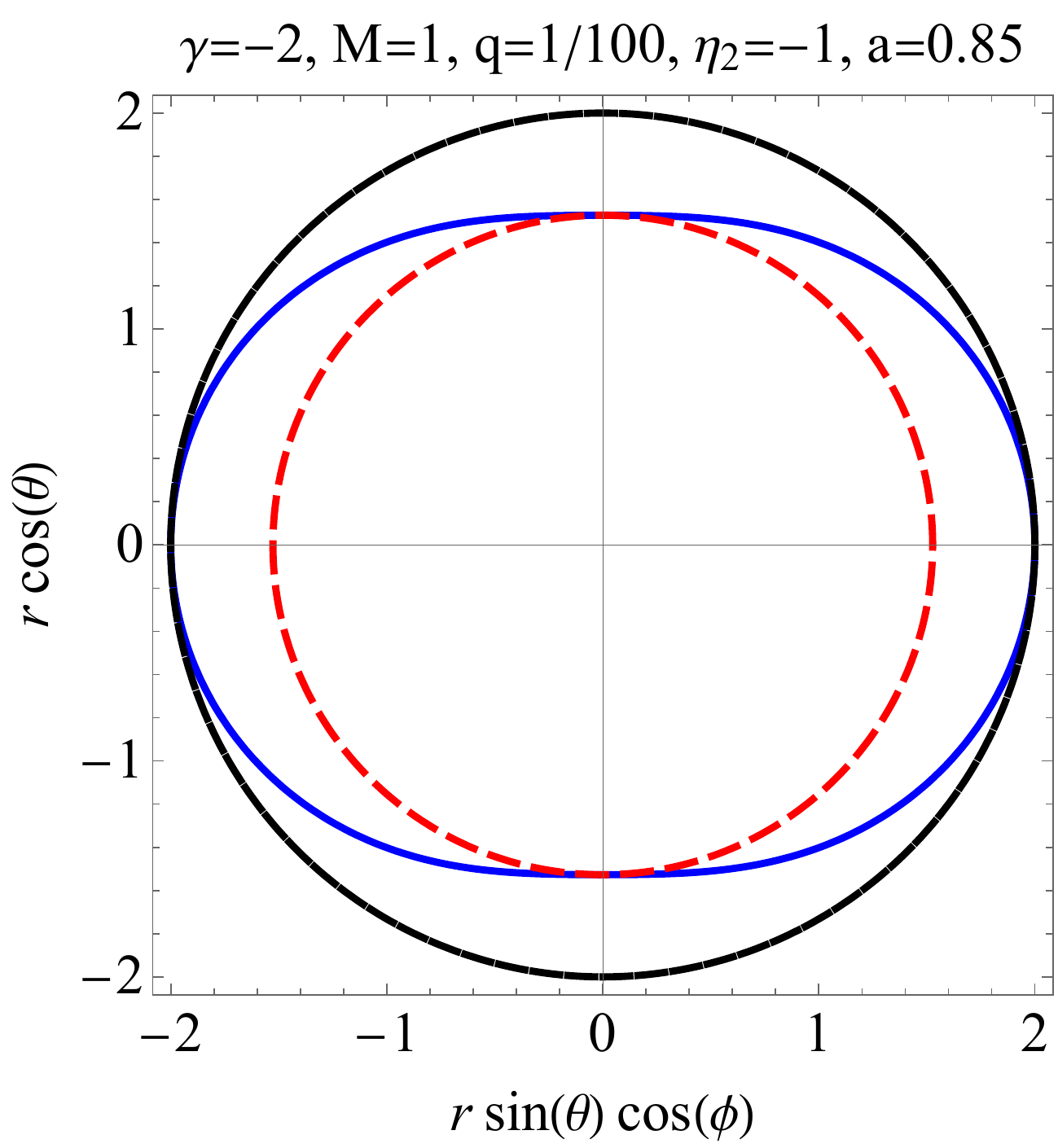}
    \includegraphics[width=0.49\linewidth]{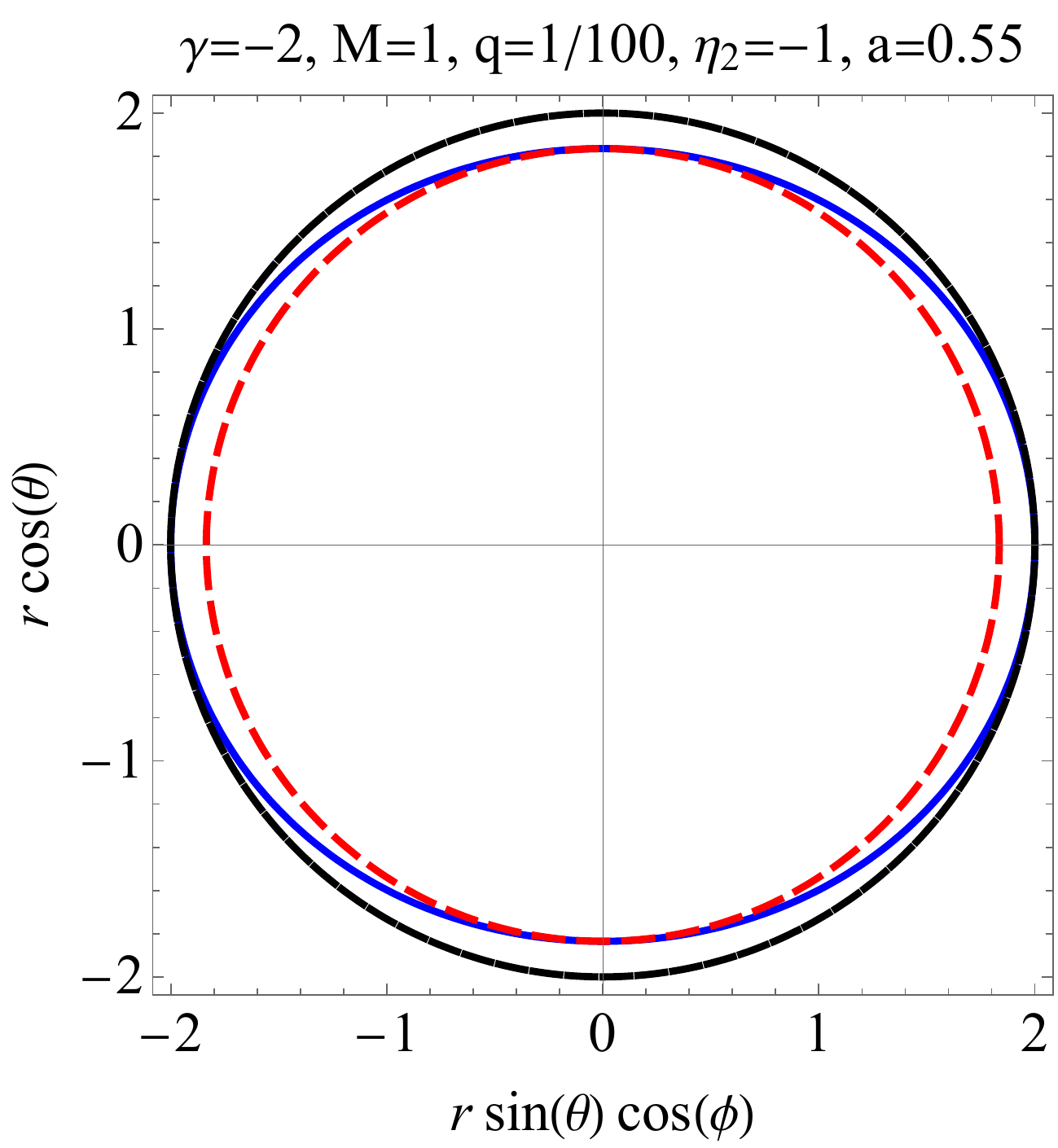}
\caption{\footnotesize{Shapes of the outer horizon (red) and outer ergo-sphere (blue). The black circle is the horizon of the phantom static black hole corresponding to $a=0$.}} \label{Figes}
\end{figure}
\begin{figure}
	\includegraphics[width=0.49\linewidth]{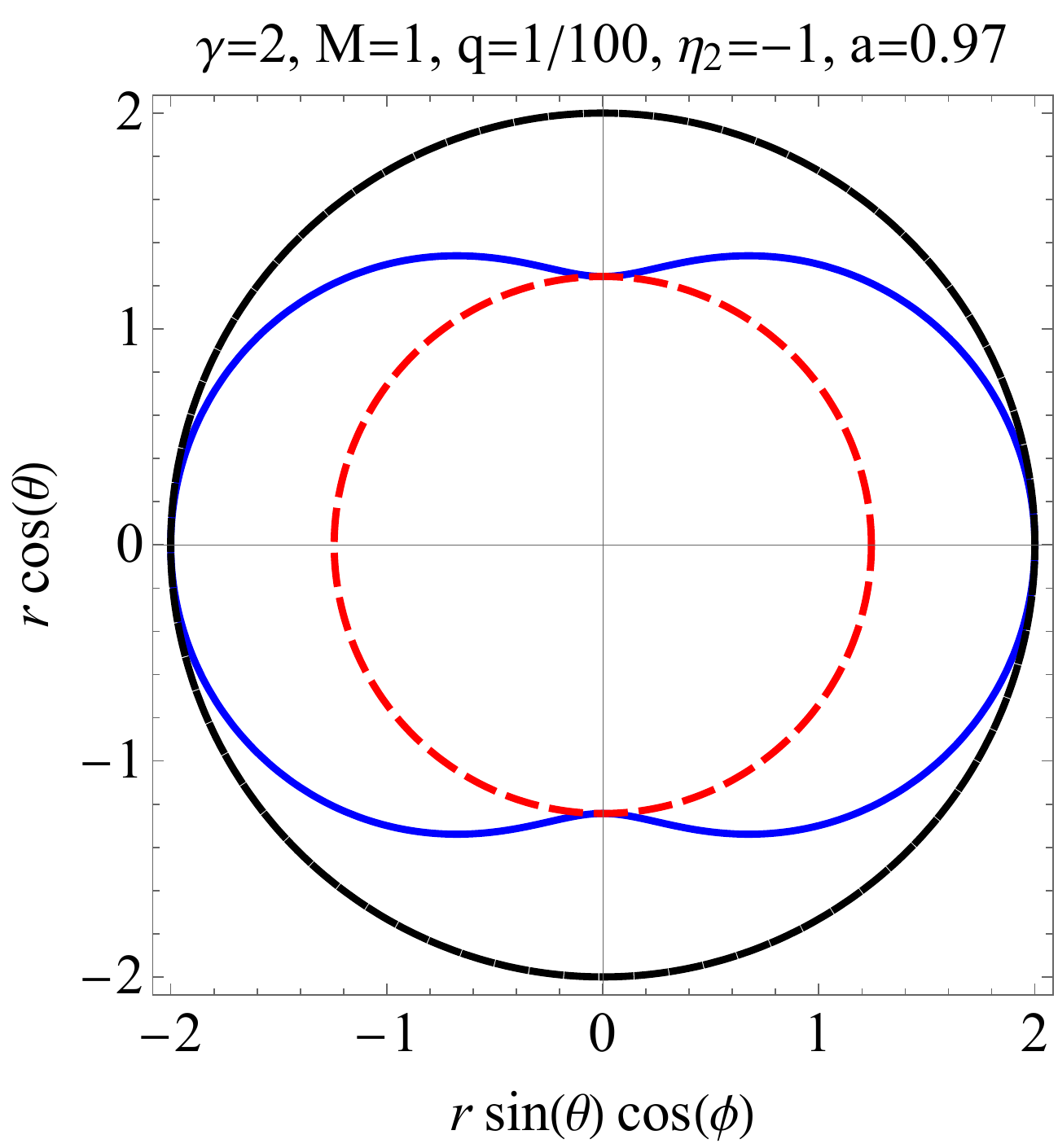}
	\includegraphics[width=0.49\linewidth]{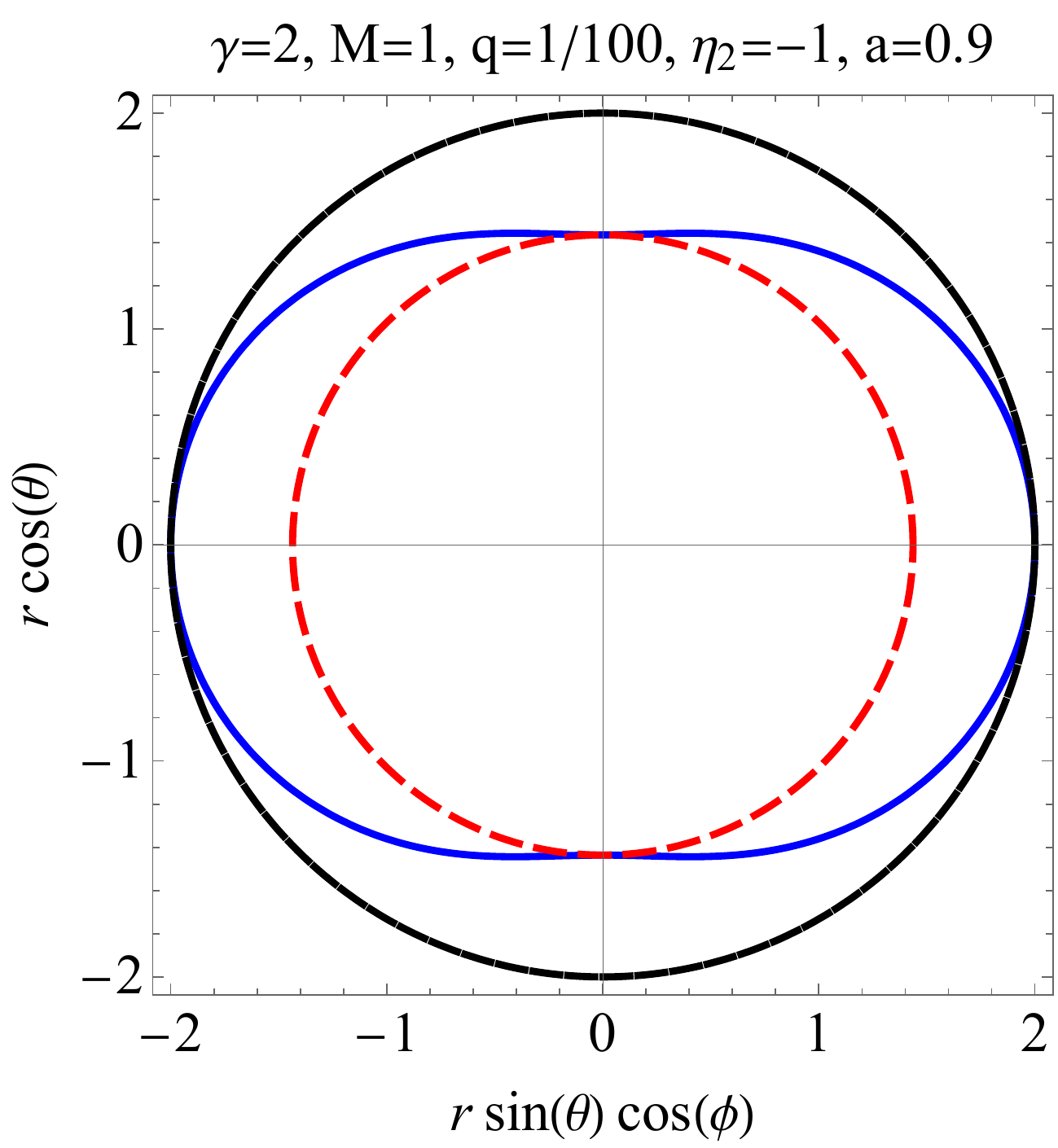}\\
	\includegraphics[width=0.49\linewidth]{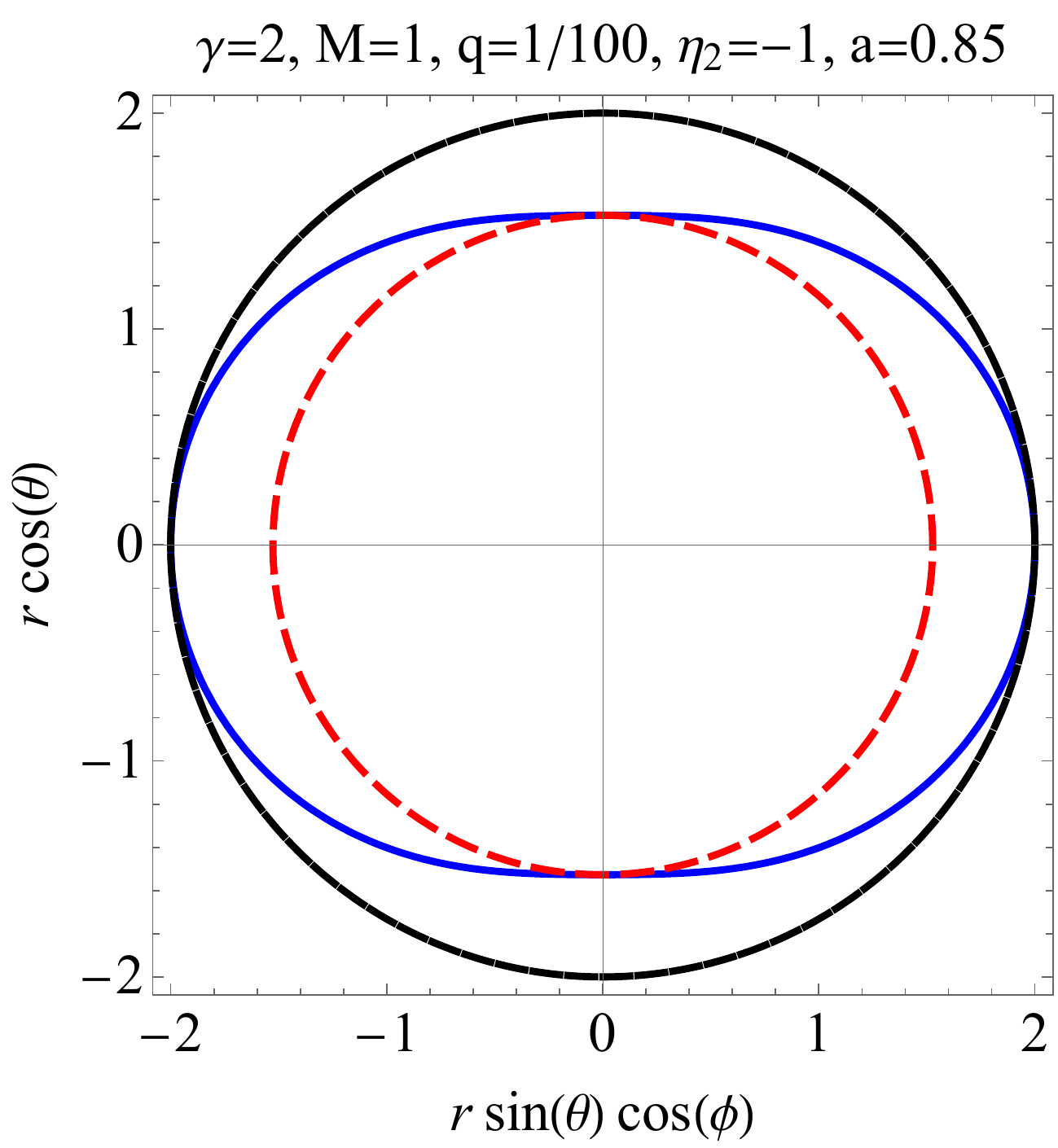}
	\includegraphics[width=0.49\linewidth]{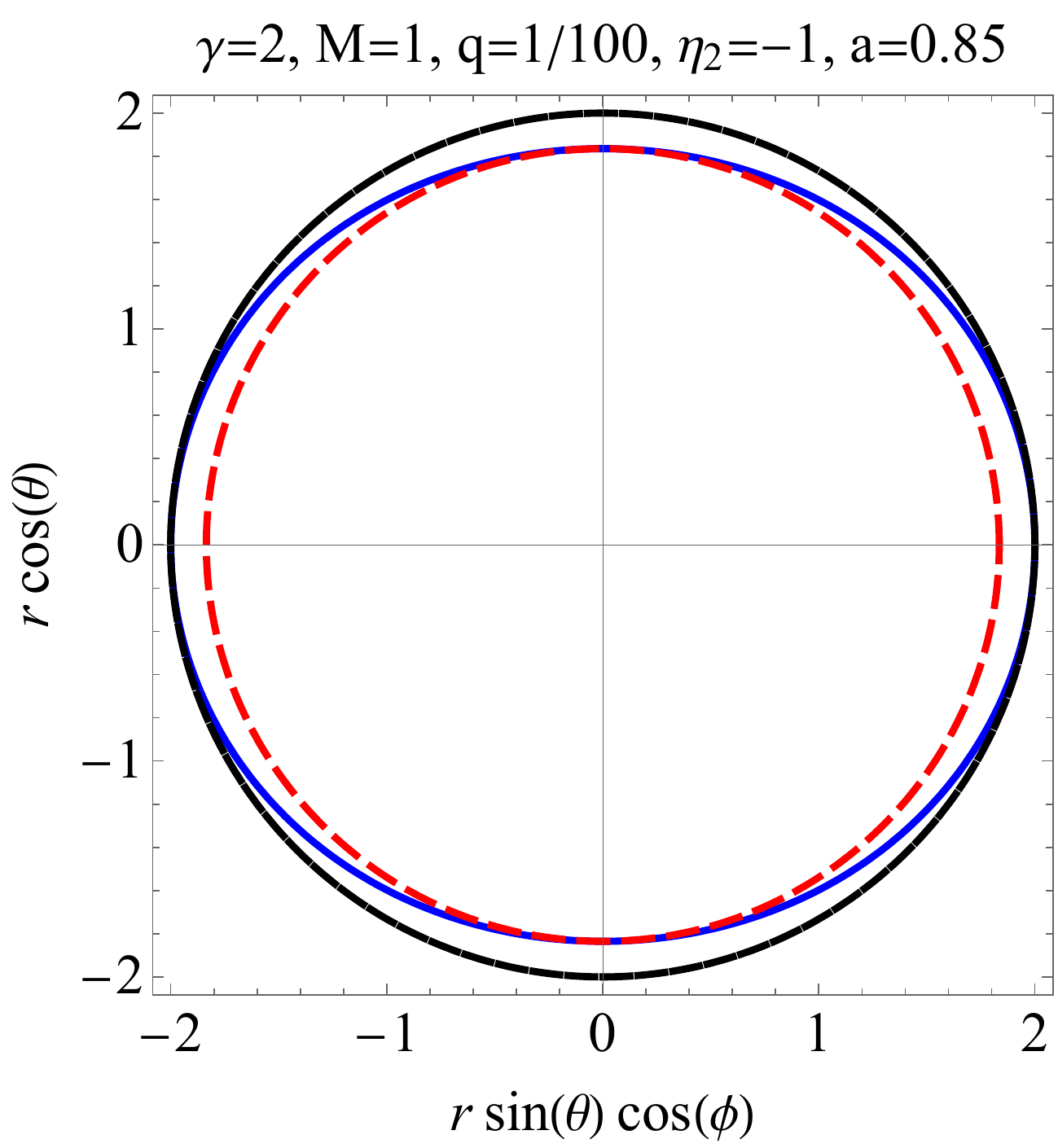}
	\caption{\footnotesize{Shapes of the outer horizon (red) and outer ergo-sphere (blue). The black circle is the horizon of the phantom static black hole corresponding to $a=0$.}} \label{Figes2}
\end{figure}
\begin{figure}
	\includegraphics[width=0.75\linewidth]{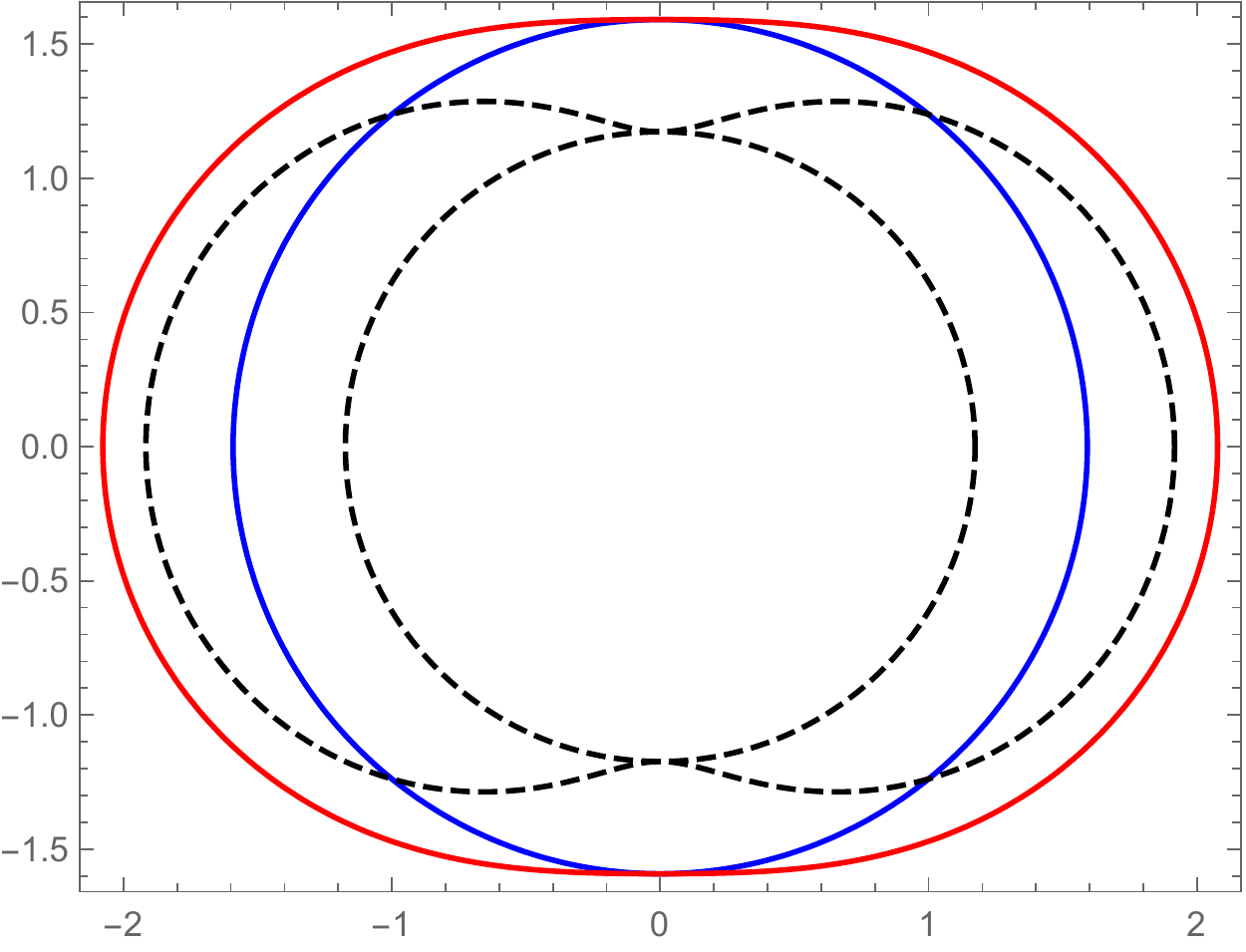}
	\caption{\footnotesize{Solid line shows the outer horizon and ergosphere for a phantom Kerr--Newman black hole while dashed lines show the outer horizon and ergosphere for normal Kerr--Newman black hole.}} \label{Figes1}
\end{figure}

\paragraph{\textbf{Ergosurfaces.}}
The the inner and outer ergo-surfaces are the two-dimensional surfaces $g_{tt}=0$ yielding
\begin{eqnarray}\nonumber
r^2-(r_1+r_2)r+r_1r_2+a^2\cos^2\theta=0.
\end{eqnarray}
The roots of this equation are as follows
\begin{equation}
r_{e_{\pm}}=\frac{1}{2}\left[\left(r_1+r_2\right)\pm\sqrt{\left(r_1-r_2\right)^2-4a^2\cos^2\theta}\right].
\end{equation}
Note that for $\theta=0 \text{ or } \pi$ the ergo-surfaces coincide with the event horizons. This property is not valid only for~\eqref{metric}; rather, it applies to all rotating metrics~\eqref{B} for all $f$, $g$, $h$, and $H$.

Figures (\ref{Figes}), (\ref{Figes2}) and (\ref{Figes1}) show the behavior of the outer horizon and outer ergosphere for different values of the parameters. We can deduce from here that the shape of the outer horizon and outer ergosphere changes with $a$. As $a\rightarrow{0}$ the horizon and ergosphere will coincide with the horizon of static EMD black hole.

\paragraph{\textbf{Singularities.}\label{parasing}}
Mathematically, the singularity in a black hole solution can be interpreted from analyzing the Kretschmann scalar $K$. This scalar, when tends to infinity, indicates the presence of curvature singularity inside the black hole. For our metric the Kretschmann scalar is
\begin{equation}
K=\mathcal{R}^{\mu\nu\rho\varsigma}\mathcal{R}_{\mu\nu\rho\varsigma}
=\frac{\mathcal{Q}}{H^6}=\frac{\mathcal{Q}}{(h+a^2\cos^2\theta)^6}=\frac{\mathcal{Q}}{\big(r^{1+\gamma}(r-r_2)^{1-\gamma}+a^2\cos^2\theta\big)^6}.
\end{equation}
where $\mathcal{Q}$ is a sizable function of ($r,\cos\theta$)  which, depending on the value of $\gamma$, may vanish if we set $h=0$ and $\cos^2\theta=0$. 

For the case $\gamma=0$ we have
\begin{equation}\label{KS0}
K=\frac{\mathcal{Q}}{\left(r^2-r_2r+a^2\cos^2\theta\right)^6},
\end{equation}
where $r_1$ and $r_2$ are given by~\eqref{hor2}. We may distinguish three cases: (1) the cosh solutions~\eqref{cd3} corresponding to $\eta_2=-1$ (always charged), (2) the sinh solutions~\eqref{cd4} corresponding to $\eta_2=+1$ (always charged), and (3) the uncharged solutions. In Ref~\cite{G} it was shown that $r=0$ is a singularity for the cosh static solutions and $r=r_2$ is a singularity for the sinh static solutions (because of the order relation~\eqref{cd2}, $r=0$ is behind the singularity at $r=r_2$ and thus is exluded from the spacetime of the solution). For the rotating cosh solutions we have
\begin{equation}
\lim_{r\to 0}K\to \frac{q^8 (a^4+27 q^4-10 a^2 q^2)}{a^{12} M^4 \cos ^{12}\theta },
\end{equation}
which diverges as $\theta\to \pi/2$. For the rotating sinh solutions we have
\begin{equation}
\lim_{r\to r_2}K\to \frac{q^8}{a^{8} M^4 \cos ^{12}\theta },
\end{equation}
which also diverges as $\theta\to \pi/2$. Thus, we conclude that the Kretschman scalar~\eqref{KS0} has a generic (for all $M$, $q$, and $a$) singularity located at:
\begin{itemize}
  \item {$r=0$ and $\theta=\pi/2$, \text{for the cosh solutions corresponding to }$\eta_2=-1$},
  \item {$r=r_2$ and $\theta=\pi/2$, \text{for the sinh solutions corresponding to }$\eta_2=+1$}.
\end{itemize}
These conclusions concern only the case $\gamma=0$.

For the rotating uncharged solutions $q=0$ we have $r_2=0$~\eqref{hor}. In this case $f$~\eqref{fg} and $h$~\eqref{hr} no longer depend on $\gamma$ and they reduce to their Schwarzschild forms and the rotating solution reduces to the Kerr black hole with
\begin{equation}
K =\frac{48 M^2 (r^6-15 a^2 r^4 \cos ^2\theta +15 a^4 r^2 \cos ^4\theta -a^6 \cos ^6\theta )}{(r^2+a^2 \cos ^2\theta )^6},\qquad(\text{for all }\gamma),
\end{equation}
and this goes to infinity as $r\to 0$ and $\theta\to \pi/2$. Thus, for all $\gamma$ and $q=0$ the Kretschman scalar~\eqref{KS0} has a generic (for all $M$, $a$, and $\gamma$) singularity located at:
\begin{itemize}
	\item {$r=0$ and $\theta=\pi/2$, \text{for the uncharged solutions}}.
\end{itemize}

\paragraph{\textbf{Area and Entropy.}}
The area of the horizon $r_{+}$ for the RPBH is given by
\begin{eqnarray}\nonumber
A=\int_{0}^{2\pi}\int_{0}^{\pi}\sqrt{g_{\theta\theta}g_{\phi\phi}}\Big|_{r=r_+}\,d\theta{d\phi}=4\pi\Big(r_+^{1+\gamma}(r_+-r_2)^{1-\gamma}+a^2\Big).
\end{eqnarray}
The entropy is then given by
\begin{eqnarray}\nonumber
S=\frac{A}{4}=\pi\Big(r_+^{1+\gamma}(r_+-r_2)^{1-\gamma}+a^2\Big).
\end{eqnarray}
For $a=0$, the area and entropy of our black hole solution matches with its static version \cite{thermo}. For $\gamma=1$ and $q=0$ we recover the area and entropy expressions of the Kerr black hole.

\section{Equatorial geodesics\label{secgeod}}
We intend to compute the null and timelike geodesics in the equatorial plane for our black hole solution. With the assumption that $\theta=\pi/2$ and $\dot{\theta}=0$, the Lagrangian for the metric (\ref{metric}) takes the form
\begin{align}
&2\mathcal{L}=\left(1-\frac{\sigma}{h}\right)\dot{t}^2-\frac{h}{\Delta}\dot{r}^2+\frac{2a\sigma}{h}\dot{t}\dot{\phi}-\left(h+a^2+\frac{a^2\sigma}{h}\right)\dot{\phi}^2,\nonumber\\
&2\mathcal{L}=f\dot{t}^2-\frac{h}{\Delta}\dot{r}^2+2a(1-f)\dot{t}\dot{\phi}-\Big(h+a^2(2-f)\Big)\dot{\phi}^2,
\end{align}
where we used $\sigma/h=1-f$~\eqref{Ds2}. From this Lagrangian, the generalized momenta are computed as
\begin{eqnarray}
p_t &=&f\dot{t}+a(1-f)\dot{\phi}=E,\\
p_\phi &=&a(1-f)\dot{t}-\Big(h+a^2(2-f)\Big)\dot{\phi}=-L,\\
p_r&=&-\frac{h}{\Delta}\dot{r}.
\end{eqnarray}
where a dot denotes differentiation with respect to the affine parameter $\tau$. Note that the Lagrangian does not dependent on $t$ and $\phi$ which yields the conservation of $p_t$ and $p_\phi$. This also depicts the stationary and axisymmetric nature of our metric.

The Hamiltonian is given by
\begin{equation}
\mathcal{H}=p_t\dot{t}+p_r\dot{r}+p_\phi\dot{\phi}-\mathcal{L}.
\end{equation}
It reduces to
\begin{equation}
2\mathcal{H}=f\dot{t}^2+2a(1-f)\dot{t}\dot{\phi}-\Big(h+a^2(2-f)\Big)\dot{\phi}^2-\frac{h}{\Delta}\dot{r}^2,
\end{equation}
\begin{equation}\label{h}
2\mathcal{H}=E\dot{t}-L\dot{\phi}-\frac{h}{\Delta}\dot{r}^2=\delta=\text{constant},
\end{equation}
\begin{eqnarray}\label{t}
\dot{t}&=&\frac{1}{\Delta}\left[\Big(h+a^2(2-f)\Big)E-a(1-f)L\right],\\\label{phi}
\dot{\phi}&=&\frac{1}{\Delta}\left[a(1-f)E+fL\right].
\end{eqnarray}
Inserting Eq.~(\ref{t}) and Eq.~(\ref{phi}) in Eq.~(\ref{h}), we obtain
\begin{equation}\label{radial}
h\dot{r}^2=hE^2+(1-f)\left(aE-L\right)^2+(a^2E^2-L^2)-\delta\Delta,
\end{equation}
which reduces to Eqs.~(53)-(54) of Ref.~\cite{A} in the case $h=r^2$. 

The separability of the Hamilton--Jacobi equation for neutral particles (for all $\theta$ and $h=r^2$) has been performed in Ref.~\cite{A} [see Eqs.~(44)-(54) of Ref.~\cite{A}].

\subsection{Null geodesics}
We can investigate the null geodesics of the RPBH on letting $\delta=0$. The radial equation (\ref{radial}) thus takes the form
\begin{equation}\label{radial1}
h\dot{r}^2=hE^2+(1-f)\left(aE-L\right)^2+(a^2E^2-L^2).
\end{equation}
Let $D=L/E$ be the impact parameter. Two cases may arise here: either $D=a$ or $D\neq{a}$. The former case is a particular case. To analyze it, let us consider $L=aE$ in Eqs.~(\ref{t}), (\ref{phi}) and (\ref{radial1}), we then obtain
\begin{eqnarray}\label{t1}
\dot{t}&=&\frac{h+a^2}{\Delta}E,\\\label{phi1}
\dot{\phi}&=&\frac{aE}{\Delta},\\\label{radial2}
\dot{r}&=&\pm{E}.
\end{eqnarray}
Note that on the horizon ($\Delta=0$), $\dot{t}$ and $\dot{\phi}$ are singular, which means that the $t$ and $\phi$ coordinates are unable to portray the trajectory of photons with respect to a co-moving observer \cite{hobson}. In Eq.~(\ref{radial2}), the plus/minus sign indicates outgoing/ingoing photons.

Now if we only consider the outgoing photons $(\dot{r}=+E)$ then the differentials of $t$ and $\phi$ with respect to $r$ are computed as
\begin{eqnarray}\label{tr}
\frac{dt}{dr}&=&\frac{h+a^2}{\Delta},\\\label{phir}
\frac{d\phi}{dr}&=&\frac{a}{\Delta},
\end{eqnarray}
leading to the solutions
\begin{eqnarray}\label{tpm}
{t}&=&\int\frac{r^{1+\gamma}(r-r_2)^{1-\gamma}}{(r-r_+)(r-r_-)}dr+\frac{a^2}{(r-r_-)}\ln{\bigg|\frac{r-r_+}{r-r_-}\bigg|},\\\label{phipm}
\phi&=&\frac{a}{r_+-r_-}\Bigg[\ln{\bigg|\frac{r}{r_+}-1\bigg|}-\ln{\bigg|\frac{r}{r_-}-1\bigg|}\Bigg].
\end{eqnarray}
In the same way, choosing $\dot{r}=-E$ will lead to solutions for incoming photons, which are obtained from the above equations upon replacing $t$ by $-t$ and $\phi$ by $-\phi$.

When $D\neq{a}$ we may obtain circular orbits for photons. To investigate this case consider an impact parameter $D=D_\text{c}=L_\text{c}/E_\text{c}$, where {\textquoteleft c\textquoteright} stands for circular orbit. For such orbits we have $r=r_\text{c}$ and $\dot{r}=0$. Thus the radial equation and its derivative are
\begin{eqnarray}\label{radial3}
h_\text{c}+(1-f_\text{c})\left(a-D_{\text{c}}\right)^2+\left(a^2-D_\text{c}^2\right)=0
\end{eqnarray}
and
\begin{eqnarray}\label{deri}
h'_\text{c}-f'_\text{c}(a-D_\text{c})^2=0
\end{eqnarray}
where $h'_\text{c}$ denotes $h'\big|_{r=r\text{c}}$ and we attach the same meaning to $f'_\text{c}$ and to further similar notation. This yields
\begin{align}
\label{y1}&(a-D_\text{c})^2=\frac{h'_\text{c}}{f'_\text{c}},\\
\label{y2}&D_c= \sqrt{\frac{h'_\text{c}}{f'_\text{c}}}\mp a,
\end{align}
where in the second line $a\geq 0$ and throughout this paper the upper/lower sign corresponds to retrograde/prograde orbits\footnote{If only one sign appears in front of $a$, as in~\eqref{radial3}, then for retrograde orbits $a$ is negative and for prograde orbits $a$ is positive. Now, if two signs appear in front of $a$ as in~\eqref{y2}, then the upper sign is for retrograde orbits and the lower sign is for prograde orbits with $a\geq 0$.}. Substituting~\eqref{y1} into~\eqref{radial3} we obtain
\begin{equation}\label{y4}
D_c^2=h_c+(1-f_c)~\frac{h'_\text{c}}{f'_\text{c}}+a^2.
\end{equation}
Now, squaring~\eqref{y2} and eliminating $D_c^2$ from~\eqref{y4} we arrive at the constraint equation
\begin{equation}\label{CPO}
1-\frac{(\ln h)'_\text{c}}{(\ln f)'_\text{c}}=\mp \dfrac{2a}{h_\text{c}}~\sqrt{\frac{h'_\text{c}}{f'_\text{c}}}.
\end{equation}

Equation~\eqref{CPO} is easily brought to
\begin{equation}\label{isco}
1-\frac{(\ln h)'_\text{c}}{(\ln f)'_\text{c}}=\mp \dfrac{2a}{\sqrt{f_\text{c}h_\text{c}}}~\sqrt{\frac{(\ln h)'_\text{c}}{(\ln f)'_\text{c}}}.
\end{equation}
This is valid for all forms of $f(r)$ and $h(r)$. The largest real positive root is the radius $r_{\text{ph}}$ of the unstable photon circular orbit.

For our RPBH solution~\eqref{1}, $fh=(r-r_1)(r-r_2)$ is independent of $\gamma$ and Eq.~\eqref{isco} takes the form
\begin{equation}\label{ip}
2r^2-[6M+(1-\gamma)r_2]r+4\eta_2q^2=\pm 2a \sqrt{\frac{(2Mr-2\eta_2q^2)[2r-(1+\gamma)r_2]}{r-r_2}},
\end{equation}
where we have omitted the subscript \textquoteleft c\textquoteright. Here we have used~\eqref{hor}: $r_1+\gamma r_2=2M$ and $(1+\gamma)r_1r_2=2\eta_2q^2$. Note that the photon circular orbit exists only if
\begin{equation}\label{icond}
r^2f'h'=\frac{(2Mr-2\eta_2q^2)[2r-(1+\gamma)r_2]}{r-r_2}\geq 0\quad\text{and}\quad r\geq r_+, 
\end{equation}
where $r_+$ is given by~\eqref{horizon}. This is the same as $h'/f'\geq 0$~\eqref{y1} and $r\geq r_+$.

\begin{figure*}
	\includegraphics[width=0.49\linewidth]{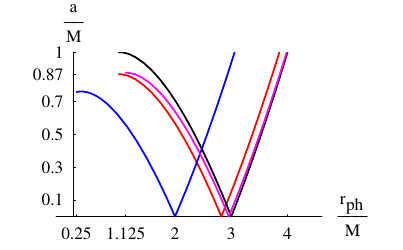} \includegraphics[width=0.49\linewidth]{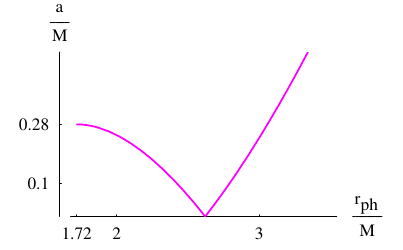}
	\caption{\footnotesize{Photon circular orbit $r_{\text{ph}}/M$ versus $a/M$ for $\eta_2=+1$. Each plot has a {\textquoteleft V\textquoteright} form where the left branch corresponds to prograde orbits and the right branch corresponds to retrograde orbits (with larger values of $r_{\text{ph}}$). The extremal RPBH corresponds to $a_{\text{ext}}=(r_1-r_2)/2$~\eqref{horizon}, the value of which is generally different from $M$ except for the uncharged solutions. For $a=a_{\text{ext}}$, $r_{\text{ph}}$ has its minimum value, $(r_1+r_2)/2$, which depends on the parameters of the nonrotating BH. Similarly, the value of $r_{\text{ph}}$ for $a=0$ depends on the parameters of the BH.\\ 
			Left Plot ($q^2<M^2$): Black Plot -- Uncharged RPBHs including the Kerr solution, Red Plot -- $q^2/M^2=0.25$ and $\gamma=1$ (Kerr--Newman), Blue Plot -- $q^2/M^2=0.25$ and $\gamma=-2$, Magenta Plot -- $q^2/M^2=0.25$ and $\gamma=0$.\\
			Right Plot ($q^2>M^2$): $q^2/M^2=1.44$ and $\gamma=0$. Here $a_{\text{ext}}=0.28M$ and the minimum $r_{\text{ph}}$ is $1.72M$.}} \label{iphoton2}
\end{figure*}
\begin{figure*}
	\includegraphics[width=0.49\linewidth]{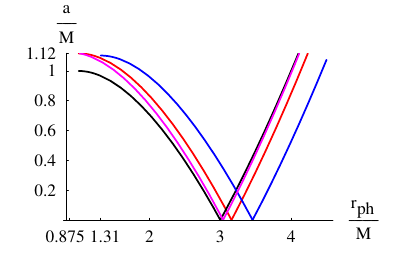} \includegraphics[width=0.49\linewidth]{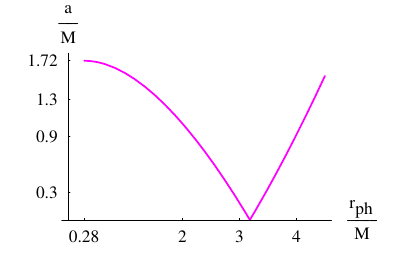}
	\caption{\footnotesize{Photon circular orbit $r_{\text{ph}}/M$ versus $a/M$ for $\eta_2=-1$. Each plot has a {\textquoteleft V\textquoteright} form where the left branch corresponds to prograde orbits and the right branch corresponds to retrograde orbits (with larger values of $r_{\text{ph}}$). The extremal RPBH corresponds to $a_{\text{ext}}=(r_1-r_2)/2$~\eqref{horizon}, the value of which is generally different from $M$ except for the uncharged solutions. For $a=a_{\text{ext}}$, $r_{\text{ph}}$ has its minimum value, $(r_1+r_2)/2$, which depends on the parameters of the nonrotating BH. Similarly, the value of $r_{\text{ph}}$ for $a=0$ depends on the parameters of the BH.\\ 
			Left Plot ($q^2<M^2$): Black Plot -- Uncharged RPBHs including the Kerr solution, Red Plot -- $q^2/M^2=0.25$ and $\gamma=1$ (phantom Kerr--Newman), Blue Plot -- $q^2/M^2=0.25$ and $\gamma=-2$, Magenta Plot -- $q^2/M^2=0.25$ and $\gamma=0$.\\
			Right Plot ($q^2>M^2$): $q^2/M^2=1.44$ and $\gamma=0$. Here $a_{\text{ext}}=1.72M$ and the minimum $r_{\text{ph}}$ is $0.28M$.}} \label{iphoton}
\end{figure*}

We can already draw the following conclusion: All the uncharged rotating normal or phantom black holes [$q=0\Rightarrow r_2=0$~\eqref{hor}] have the same photon circular orbit for all values of $\gamma$. This is the case because the metric of uncharged RPBHs reduces to the Kerr metric (in Sec.~\ref{parasing} we have noticed that $f$~\eqref{fg} and $h$~\eqref{hr} no longer depend on $\gamma$ once we take $r_2=0$ and they reduce to their Schwarzschild forms). Now, setting $q=0$, Eq.~\eqref{ip} reduces to its coreesponding Kerr equation
\begin{equation}\label{ip2}
(r-3M)\sqrt{r}=\pm 2a\sqrt{M}.
\end{equation}
This can be solved on setting $\sqrt{r}=x$ and using the Weierstrass polynomial as shown in Appendix A of Ref.~\cite{acc}. Its largest real root is given by~\cite{Teukolsky,Chandra}
\begin{equation}\label{ip3}
r_{\text{ph}}=2M\Big\{1+\cos\Big[\frac{2}{3}\arccos\Big(\pm \frac{a}{M}\Big)\Big]\Big\}.
\end{equation}

In the charged case $r_{\text{ph}}$ is a function of ($M,q^2,a,\text{sgn}(a),\gamma,\eta_2$). Eq.~\eqref{ip} can be brought to an algebraic equation of the form $r^5+\cdots=0$ upon squaring both sides. The latter is solved with the constraint $2r^2-[6M+(1-\gamma)r_2]r+4\eta_2q^2>0$ for retrograde orbits and the constraint $2r^2-[6M+(1-\gamma)r_2]r+4\eta_2q^2<0$ for prograde orbits. Since the largest root of the equation,
\begin{equation}\label{ip4}
2r^2-[6M+(1-\gamma)r_2]r+4\eta_2q^2=0,
\end{equation}
provides the photon circular orbit for static (nonrotating) solutions by
\begin{equation}\label{ip5}
r_{\text{static}}=\frac{6M+(1-\gamma)r_2+\sqrt{[6M+(1-\gamma)r_2]^2-32\eta_2q^2}}{4},
\end{equation}
we see that the photon circular orbit for retrograde orbits must be greater than $r_{\text{static}}$ and the photon circular orbit for prograde orbits must be smaller than $r_{\text{static}}$.

For some special values of $\gamma$, Eq.~\eqref{ip} can be brought to more simplified expression as is the case for $\gamma=0$:
\begin{equation}\label{ip6}
X^4+(\eta_2q^2-6M^2)X^2\mp 4\sqrt{2}aM^2X+2\eta_2M^2q^2=0,
\end{equation}
where $2Mr_{\text{ph}}=X^2+\eta_2q^2$ ($X\equiv\sqrt{2Mr-\eta_2q^2}$).

In Figs.~\ref{iphoton2} and ~\ref{iphoton} we depict plots of the photon circular orbit $r_{\text{ph}}/M$ versus $a/M$ for $\eta_2=+1$ and $\eta_2=-1$. Each plot has a {\textquoteleft V\textquoteright} form where the left branch corresponds to prograde orbits and the right branch corresponds to retrograde orbits (with larger values of $r_{\text{ph}}$). The extremal RPBH corresponds to $a_{\text{ext}}=(r_1-r_2)/2$~\eqref{horizon}, the value of which is generally different from $M$ except for the uncharged solutions. For $a=a_{\text{ext}}$, $r_{\text{ph}}$ has its minimum value, $(r_1+r_2)/2$, which depends on the parameters of the nonrotating BH. Similarly, the value of $r_{\text{ph}}$ for $a=0$ depends on the parameters of the BH.

\subsection{Time-like geodesics}
\begin{figure}[h]
\includegraphics[width=0.45\textwidth]{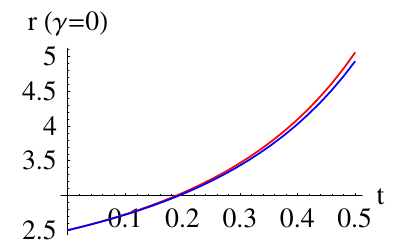}
\includegraphics[width=0.45\textwidth]{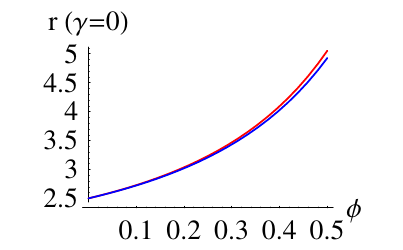}\\
\includegraphics[width=0.45\textwidth]{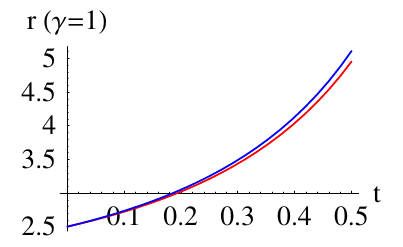}
\includegraphics[width=0.45\textwidth]{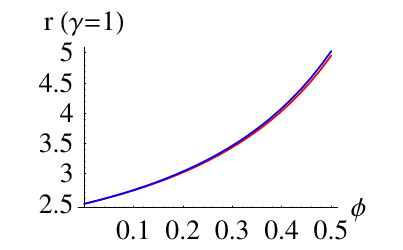}\\
\includegraphics[width=0.45\textwidth]{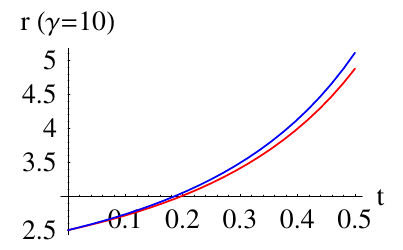}
\includegraphics[width=0.45\textwidth]{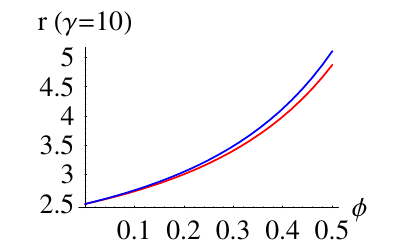}
\caption{\footnotesize{$r$ versus $t$ and $r$ versus $\phi$ time-like geodesics for $L=aE$, $E=1$, $M=1$, $q=1/2$, $a=0.82$ and different values of $\gamma$. The blue line corresponds to $\eta_2=+1$ and the red line corresponds to $\eta_2=-1$. The initial conditions are: $r(t=0)=2.5>r_+$~\eqref{horizon} and $r(\phi=0)=2.5>r_+$.}}\label{figure2}
\end{figure}

For the time-like geodesics we consider $\delta=1$. Equations for $\dot{\phi}$ and $\dot{t}$ remain the same but the radial equation~(\ref{radial}) becomes
\begin{equation}\label{timer}
h\dot{r}^2=hE^2+(1-f)\left(aE-L\right)^2+(a^2E^2-L^2)-\Delta.
\end{equation}
For the case $L=aE$, the above equation reduces to
\begin{equation}\label{TG2}
\dot{r}=\pm\sqrt{E^2-\frac{\Delta}{h}}=\pm\sqrt{\frac{1}{h}[(hE^2-r^2)+(r_1+r_2)r-r_1r_2-a^2]}\,,
\end{equation}
where $+(-)$ sign stand for outgoing (ingoing) motion and~\eqref{Ds1} has been used. The $t(r)$ ans $\phi(r)$ functions (for the outgoing geodesics) are derived using Eqs.~(\ref{t1}), (\ref{phi1}), and (\ref{TG2})
\begin{align}\label{t2}
&\frac{dr}{dt}=\frac{\Delta}{h+a^2}\sqrt{1-\frac{\Delta}{E^2h}},\\\label{phi2}
&\frac{dr}{d\phi}=\frac{\Delta}{a}\sqrt{1-\frac{\Delta}{E^2h}}.
\end{align}
These equations are numerically solved and plotted in Fig.~(\ref{figure2}) for different values of $\eta_1$~\eqref{gdom} and $\eta_2$. 

For the general case $L\neq{aE}$ Eq. (\ref{t}), Eq. (\ref{phi}) and Eq. (\ref{timer}) are written as
\begin{eqnarray}\label{t3}
\dot{t}&=&\frac{1}{\Delta}\left[\left(h+a^2\right)E+a(1-f)\left(aE-L\right)\right],\\\label{phi3}
\dot{\phi}&=&\frac{1}{\Delta}\left[(1-f)\left(aE-L\right)+L\right],\\\label{r3}
\dot{r}&=&\pm\sqrt{\frac{1}{h}\left[hE^2+(1-f)\left(aE-L\right)^2+\left(a^2E^2-L^2\right)-\Delta\right]},
\end{eqnarray}
yielding (for the outgoing geodesics)
\begin{eqnarray}\label{rt}
\frac{dr}{dt}&=&\frac{(\Delta/\sqrt{h})}{\left(h+a^2\right)E+a(1-f)\left(aE-L\right)}\sqrt{hE^2+(1-f)\left(aE-L\right)^2+\left(a^2E^2-L^2\right)-\Delta},\\\label{rphi}
\frac{dr}{d\phi}&=&\frac{(\Delta/\sqrt{h})}{(1-f)\left(aE-L\right)+L}\sqrt{hE^2+(1-f)\left(aE-L\right)^2+\left(a^2E^2-L^2\right)-\Delta}.
\end{eqnarray}
We have solved numerically these two equations and we have found that their plots are very similar to the plots shown in Fig.~(\ref{figure2}).

In this case ($L\neq aE$) we may have circular time-like orbits. In order to derive them we proceeded as in Eqs.~(\ref{radial3}) and~(\ref{deri}). The rhs of~\eqref{timer} and its derivative with respect to $r$ must vanish, this yields the following equation (we omit the subscript \textquoteleft c\textquoteright):
\begin{align}
\label{tl1}&E^2h-\Delta+E^2u^2(1-f)+E^2(2a-u)u=0,\\
\label{tl2}&E^2h'-\Delta'-E^2u^2f'=0,
\end{align}
where $u\equiv a-D$. These equations provides the values of the energy and angular momentum for a given radius by
\begin{align}
&u=\frac{a (h f'+f h')-(a^2+f h) \sqrt{f' h'}}{f^2 h'-a^2 f'},\nonumber\\
\label{tl3}&E=\bigg(\frac{f^3 (h')^2-a^2 h (f')^2-f' [f (3 a^2+f h) h'+2 a (a^2+f h) \sqrt{f' h'}]}{(f
	h'-f' h+2 a \sqrt{f' h'}) (f h'-f' h-2 a \sqrt{f' h'})}\bigg)^{1/2},\\
\label{tl4}&L=E(a-u),
\end{align}
where we have used $\Delta=fh+a^2$~\eqref{Ds1}. Notice that the existence of circular orbits is subject to the same constraint~\eqref{icond}. Another constraint is that $E$ must be real.

For all uncharged RPBH, Eqs~\eqref{tl3} and~\eqref{tl4} reduce to their known Kerr expressions
\begin{equation}
E=\frac{\sqrt{r} (r-2 M)\mp a \sqrt{M}}{\sqrt{r^3-3 M r^2\mp 2 a r \sqrt{M r}}},\qquad L= \frac{\sqrt{M} (a^2+r^2\pm 2 a \sqrt{M r})}{\sqrt{r^3-3
		M r^2\mp 2 a r \sqrt{M r}}}.
\end{equation}
To determine the last stable circular orbit (lsco) for time-like circular orbits, corresponding to the minimum of $E$, we have to differentiate the expression of $E(r)$~\eqref{tl3} and set $\partial_r E=0$ to obtain the value of $r_\text{lsco}$. However, the expression of $\partial_r E$ is sizable if $q\neq 0$ even if we restrict ourselves to some specific values of the parameters. For the uncharged RPBH ($q= 0$), $r_\text{lsco}$ is solution to the following quartic equation, which is the corresponding equation for the Kerr solution~\cite{Chandra},
\begin{equation}\label{tl5}
r^2-6Mr\mp 8a\sqrt{Mr}-3a^2=0.
\end{equation}

\begin{figure*}[h!]
\includegraphics[width=0.49\linewidth]{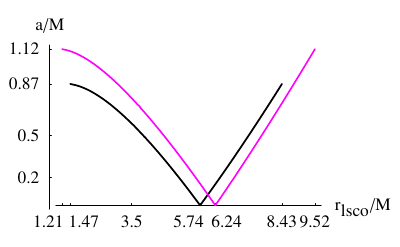} \includegraphics[width=0.49\linewidth]{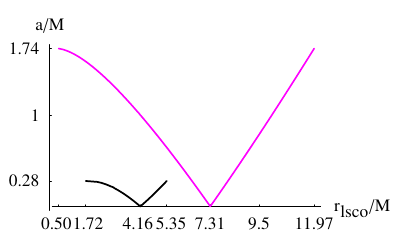}
	\caption{\footnotesize{Last stable circular orbit $r_{\text{lsco}}/M$ (the so-called innermost stable circular orbit) versus $a/M$ for normal BHs $\eta_2=+1$ (black plots) and phantom BHs $\eta_2=-1$ (magenta plots). Each plot has a {\textquoteleft V\textquoteright} form where the left branch corresponds to prograde orbits and the right branch corresponds to retrograde orbits (with larger values of $r_{\text{lsco}}$). In each plot $a/M$ runs from 0 to its upper limit $a_{\text{ext}}/M=(r_1-r_2)/(2M)$~\eqref{horizon}.\\ 
			Left Plot ($q^2<M^2$): $q^2/M^2=0.25$ and $\gamma=0$. The upper limits of $a/M$ are 0.87 for $\eta_2=+1$ and 1.12 for $\eta_2=-1$.\\
			Right Plot ($q^2>M^2$): $q^2/M^2=1.44$ and $\gamma=0$. The upper limits of $a/M$ are 0.28 for $\eta_2=+1$ and 1.74 for $\eta_2=-1$.}} \label{figlsco}
\end{figure*}

For charged RPBHs we restrict ourselves to the case $\gamma=0$, which corresponds to $\eta_1=+1$~\eqref{gdom} and
\begin{equation*}
f=1-\dfrac{2M}{r},\qquad h=r^2-\frac{\eta_2q^2}{M}\ r,
\end{equation*}
and solve numerically the equation $\partial_r E=0$ for the values of $r$ that minimize $E$. This yields the four plots shown in Fig.~\ref{figlsco} where the black plots correspond to normal BHs ($\eta_2=+1$) and the magenta plots correspond to phantom BHs ($\eta_2=-1$). It is clear from these plots that the spatial extent of $a$ and $r_{\text{lsco}}$ (in both directions) is much larger for the phantom case (magenta plots). While for $q^2<M^2$ the plots for $\eta_2=+1$ and $\eta_2=-1$ overlap, in the case $q^2>M^2$ they may not overlap as shown in the right plot of Fig.~\ref{figlsco}.

\subsection{Effective potential}
In this section we investigate the stability of the circular motion. For that end we rewrite the radial equation (\ref{radial}) in the form
\begin{eqnarray*}
	\frac{E^2-1}{2}=\frac{\dot{r}^2}{2}+V_\text{eff},
\end{eqnarray*}
where $V_\text{eff}$ stands for the effective potential. For the metric~\ref{metric} the effective potential is given by
\begin{eqnarray}\label{ep1}
	V_\text{eff}=-\frac{1-f}{2h}\left(L-aE\right)^2-\frac{\left(a^2E^2-L^2\right)-\delta\Delta}{2h}-\frac{1}{2},\qquad (L\neq{aE}),
\end{eqnarray}
where $\delta=0$ for null geodesics and $\delta=1$ for time like geodesics. Recall that circular orbits exist for $L\neq{aE}$ only. The conditions for having a circular orbit at $r=r_o$ are: $\dot{r}=0$ and $\ddot{r}=0$ at $r=r_o$. Alternatively we can say that
\begin{figure}[h!]
	\includegraphics[width=0.45\linewidth]{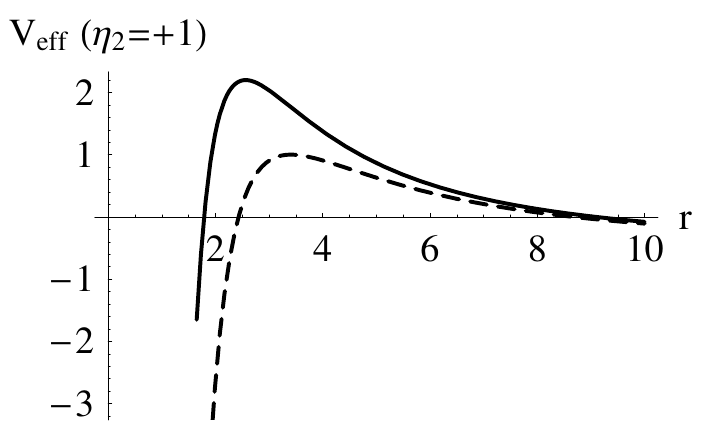}
	\includegraphics[width=0.45\linewidth]{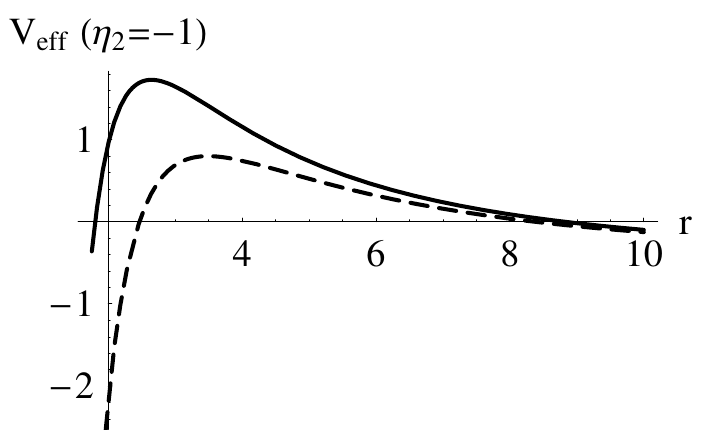}
	\includegraphics[width=0.45\linewidth]{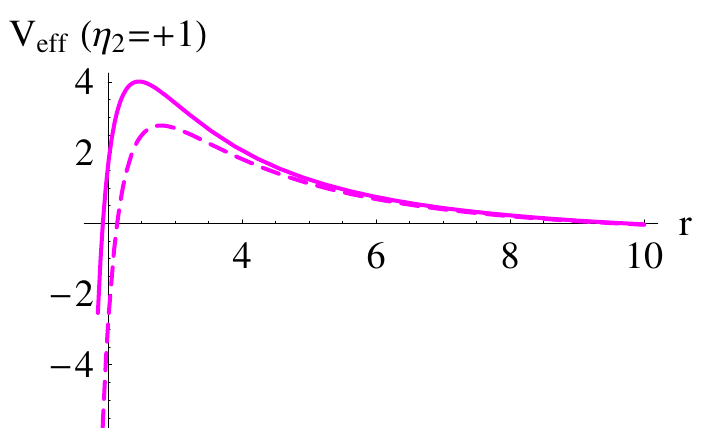}
	\includegraphics[width=0.45\linewidth]{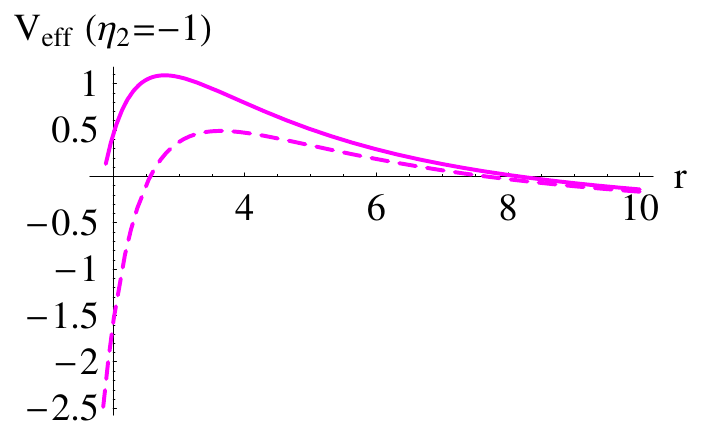}
	\caption{\footnotesize{Null geodesics: Plots of the effective potential versus $r$ for $\gamma=0$ and $L\neq{aE}$. Solid (Dashed) lines correspond to $L>0$ ($L<0$). Black (Magenta) lines correspond to $M^2<q^2$ ($M^2>q^2$).}} \label{eff3}
\end{figure}
\begin{eqnarray}
	V_\text{eff}\big|_{r=r_o}=\frac{E^2-1}{2}\quad\text{and}\quad\frac{dV_\text{eff}}{dr}\Big|_{r=r_o}=0.
\end{eqnarray}
To ensure that the circular orbit is stable, the value $(E^2-1)/2$ must be a minimum of the effective potential i.e.
\begin{eqnarray}
	\frac{d^2V_\text{eff}}{dr^2}\Big\vert_{r=r_o}>0.
\end{eqnarray}
In Figure(\ref{eff3}), the $r$-variation of the effective potential~\eqref{ep1} for null-geodesics ($\delta=0$) is depicted. We notice that the effective potential has only local maxima for fixed $E$ and different values of the angular momentum $L$. The unstable circular orbits exist at the local maxima of the effective potential. The same result is observed for photons orbits around the Schwarzschild and Kerr black holes.

\begin{figure}[h!]
	\includegraphics[width=0.49\textwidth]{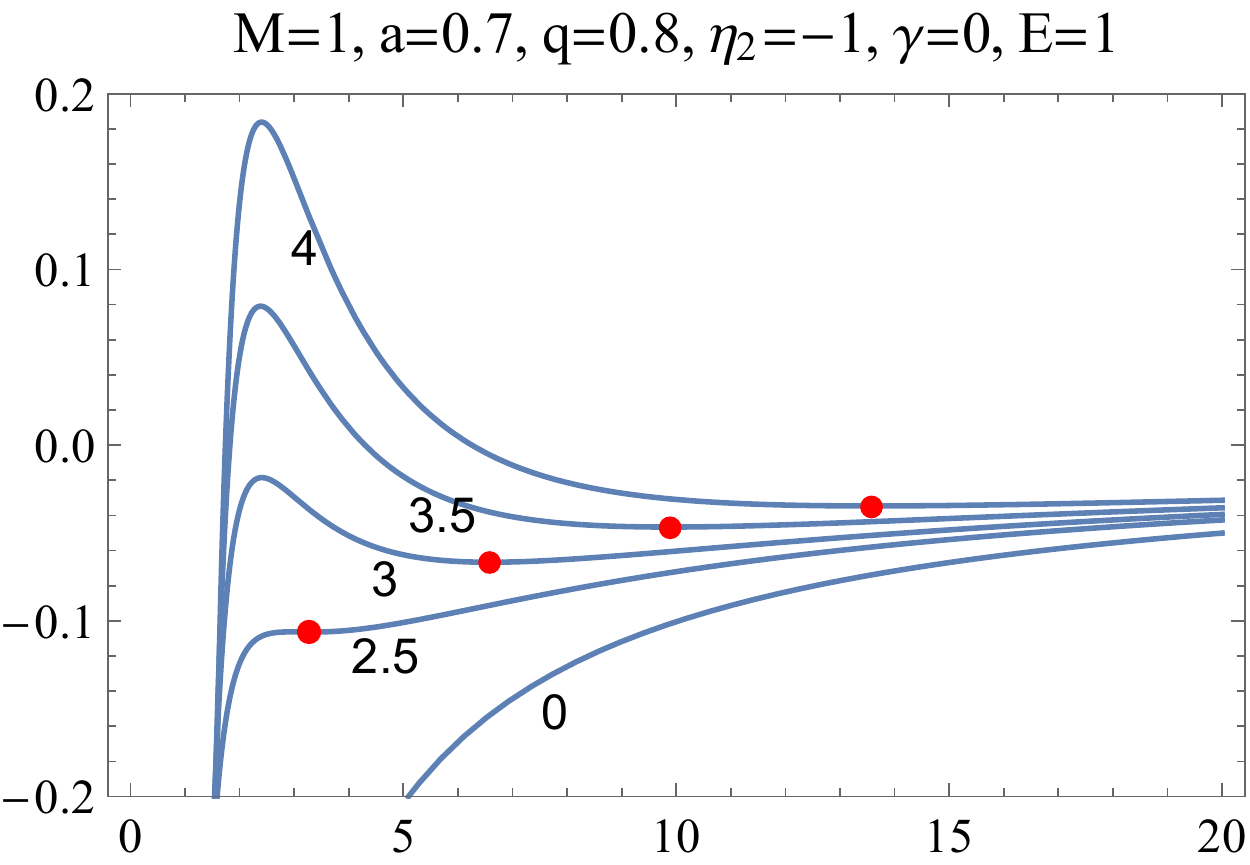}
	\includegraphics[width=0.49\textwidth]{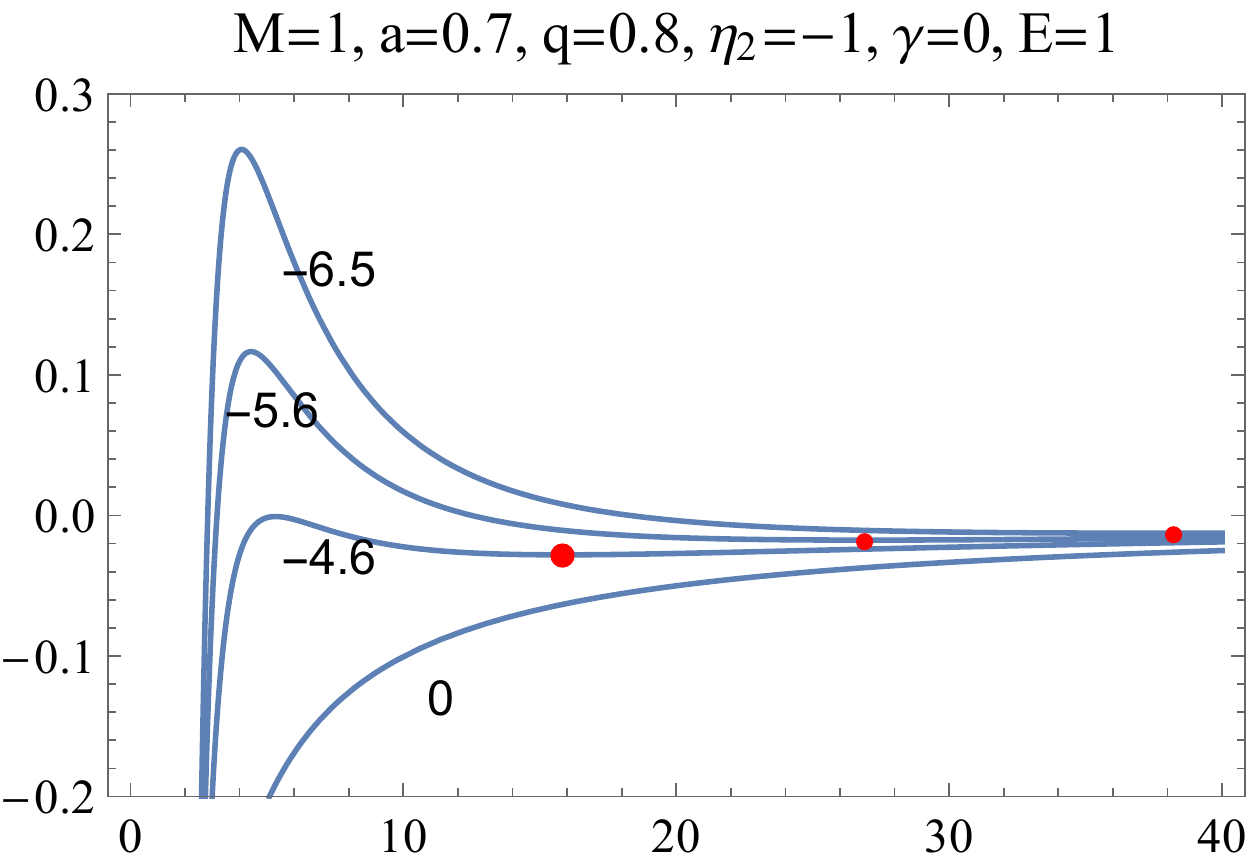}
	\caption{\footnotesize{Time-like geodesics: Plots of the effective potential versus $r$ for $\gamma=10$ and $L\neq{aE}$. The left panel corresponds to $L>0$ and the right panel corresponds to $L<0$. Radii of stable circular orbits are shown by red dots.}}\label{eff2}
\end{figure}

In Figure (\ref{eff2}), the effective potential~\eqref{ep1} for time-like geodesics ($\delta=1$) is plotted for various values of $L\ (\neq {aE})$. The local minima, which give the radii of stable circular orbits, are shown by red dots.

\section{Lense--Thirring precession\label{secLT}}

Another characteristic feature worth studying is Lense--Thirring precession, which has its importance not only in Theory of General Relativity but also in Astrophysics. Due to non-static stationary distribution of mass-energy, rotation of spacetime for instance, Einstein's theory of general relativity predicts frame dragging effects, also widely known as Lense-Thirring effect. This effect suggests that a massive object, if rotating, will cause the spacetime to distort which will further result in the precession of the orbit of a nearby particle. The precession frequency due to this rotation is known as Lense--Thirring precession frequency \cite{LT} and its co-vector is given by \cite{LTformula1,LTCM}
\begin{equation}\label{LT}
	\widetilde{\Omega}_{LT}=\frac{\varepsilon _{ijl}}{2\sqrt{-g}}\left[
	g_{ti,j}\left( \partial _{l}-\frac{g_{tl}}{g_{tt}}\partial _{t}\right) -%
	\frac{g_{ti}}{g_{tt}}g_{tt,j}\partial _{l}\right] ,
\end{equation}%
where $\varepsilon _{ijl}$ is the Levi--Civita symbols ($\varepsilon _{r\theta\phi}=+1$) and $-g=H^2\sin^2\theta$ is the determinant of the metric $g_{\mu\nu}$. Here the summation is over the spatial coordinates ($i,j,l$). The above expression suggests that the Lense--Thirring precession will not occur for the case when $g_{ti}=0$ i.e. for static black holes. We have already discussed that ergosurfaces exist when $g_{tt}=0$. Notice that (\ref{LT}) is divergent at $g_{tt}=0$ which indicates that the phenomenon of Lense--Thirring precession occurs only outside the ergoregion of the black hole.

For stationary axisymmetic spacetime the corresponding vector field with coordinates $(t,r,\theta,\phi),$ is given by%
\begin{equation*}
	\vec{\Omega}_{LT}=\frac{1}{2\sqrt{-g}}\left[ -\sqrt{-g_{rr}}\left( g_{t\phi
		,\theta }-\frac{g_{t\phi }}{g_{tt}}g_{tt,\theta }\right) \hat{r}+\sqrt{
		-g_{\theta \theta }}\left( g_{t\phi ,r}-\frac{g_{t\phi }}{g_{tt}}%
	g_{tt,r}\right) \hat{\theta}\right],
\end{equation*}%
where $\hat{r}$ and $\hat{\theta}$ are basic vectors along $r$ and $\theta $ directions. Using the metric coefficients we
arrive at
\begin{equation}
	\vec{\Omega}_{LT}=-\frac{a}{2\left(H\right) ^{3/2}\left(\Delta-a^2\sin^2\theta\right) }\left[{2\sigma \sqrt{\Delta}\cos \theta }\, \hat{r}+\sin \theta \left(
	H\sigma _{,r}-\sigma  H_{,r}\right) \hat{%
		\theta}\right],
\end{equation}
which is defined for $\Delta> 0$ only. The magnitude of the Lense--Thirring precession frequency is thus computed as
\begin{equation}
	\Omega _{LT}=\frac{a}{2(H)^{3/2}(\Delta-a^2\sin^2\theta)}\sqrt{{4\sigma ^{2}\Delta\cos ^{2}\theta }+\sin ^{2}\theta \left( H
		\sigma _{,r}-\sigma H _{,r}\right) ^{2}} .
\end{equation}
It is understood that this is evaluated for $\Delta>a^2\sin^2\theta\geq 0$, that is, outside the ergoregion where $g_{tt}\propto \Delta-a^2\sin^2\theta =fh+a^2\cos^2\theta=H-\sigma>0$. 
\begin{figure}[t]
	\centering
	\includegraphics[width=8.8cm,height=6.0cm]{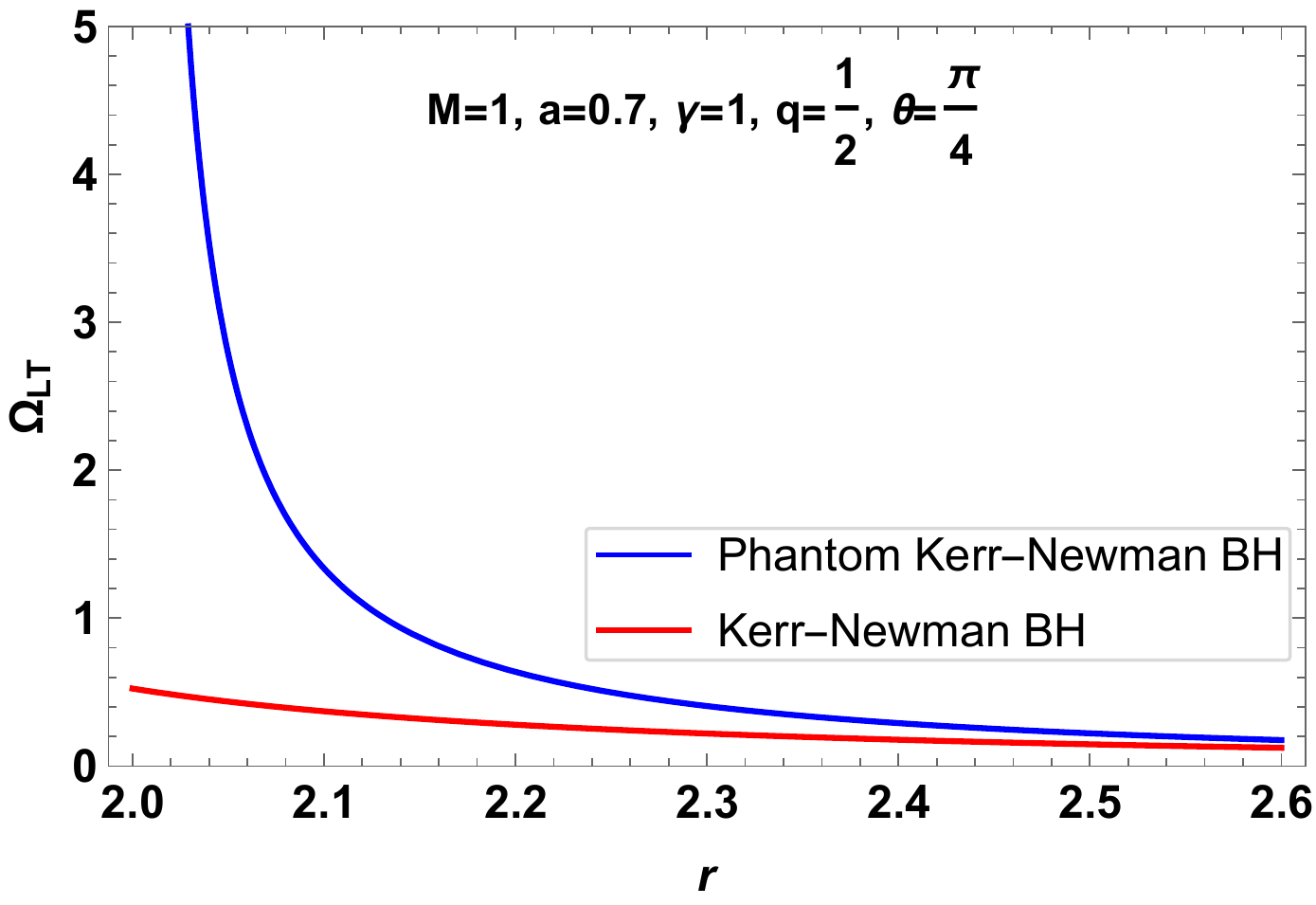}
	\caption{\footnotesize{Comparison between magnitude of spin precession frequency $\Omega_{LT}$ of normal and Phantom Kerr--Newman black holes.}}\label{LTKN}
\end{figure}
\begin{figure}[h!]
	\includegraphics[width=0.49\textwidth]{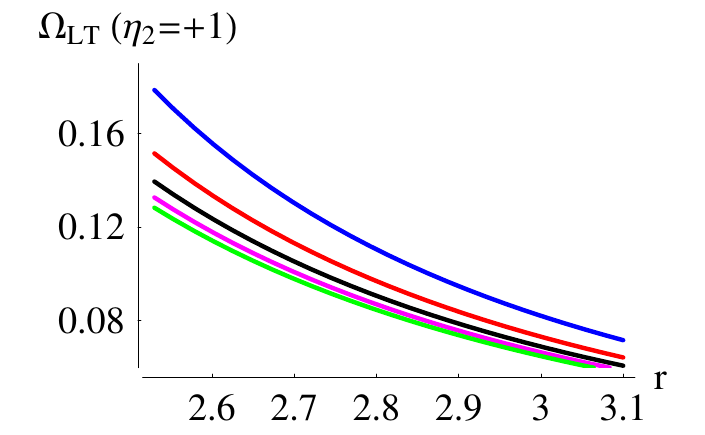}
	\includegraphics[width=0.49\textwidth]{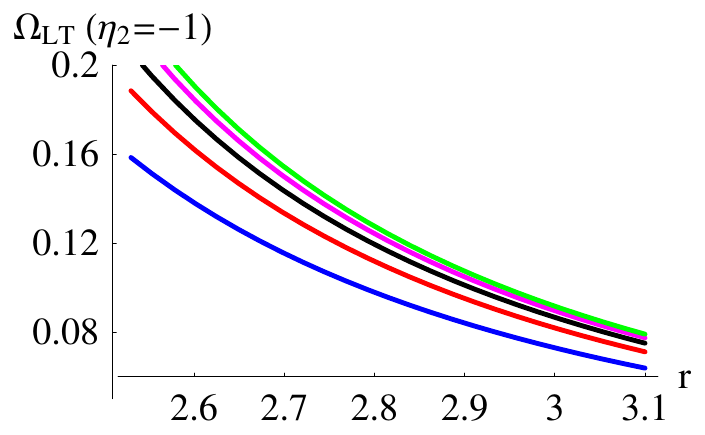}
	\caption{\footnotesize{Magnitude of the spin precession frequency $\Omega_{LT}$ versus $r$ for $M=1$, $q=1/2$, $a=0.7$, $\theta=\pi/4$, and different values of $\gamma$ and $\eta_2$. The blue plots correspond to $\gamma =0$, the red plots to $\gamma =0.5$, the black plots to normal and phantom Kerr--Newman black holes ($\gamma =1$), the magenta plots to $\gamma =1.5$, and the green plots to $\gamma =2$.}} \label{FigO1}
\end{figure}

For $\gamma =1$ and $q=0$ the Lense--Thirring precession frequency for Kerr black hole is successfully recovered \cite{LTCM}. For $\gamma=1$, $\eta_2=1$ and $q\neq{0}$, the precession frequency for Kerr--Newman black hole is obtained \cite{LTKN}.  A graphical presentation of the comparison between magnitudes of the precession frequency of normal Kerr--Newman and phantom Kerr--Newman black holes is provided in Fig.~(\ref{LTKN}), where it is shown that the presence of phantom parameter $\eta_2$ ($=-1$) enhances the Lense--Thirring effect of the black hole. For the set of parameters used in Fig.~(\ref{LTKN}) the precession frequency $\Omega_{LT}$ shows divergence at $r\approx2.0025$, for phantom Kerr--Newman black hole, while for normal Kerr--Newman case it diverges at $r\approx1.71063$, this is because, generally, the ergoregion for phantom black holes contains the ergoregion for normal black holes as shown in Fig.~\ref{Figes1}. The fact that the passage from a normal case ($\eta_2=+1$) to a phantom case ($\eta_2=-1$) enhances the Lense--Thirring effect, as emphasized in Fig.~\ref{FigO1}, remains true for all $\gamma>0$. For $\gamma=0$ the behavior is quite different; rather, it is the passage from the phantom case to the normal one that enhances the effect. From the same figure we see that $\Omega _{LT}$ decreases with $\gamma\geq 0$ if $\eta_2=+1$ and it increases if $\eta_2=-1$.
\begin{figure}[h!]
	\includegraphics[width=0.49\textwidth]{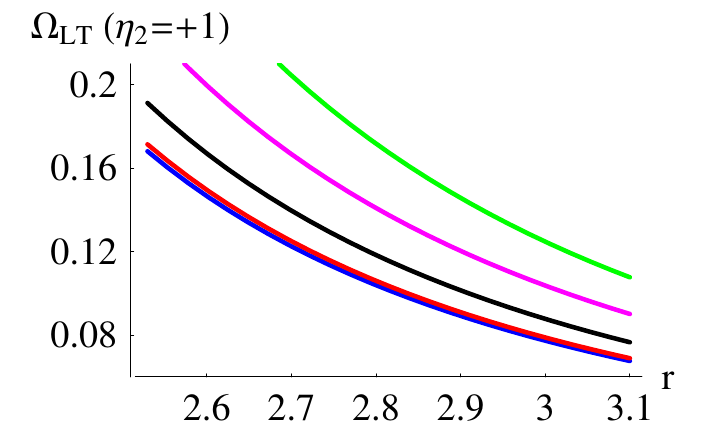}
	\includegraphics[width=0.49\textwidth]{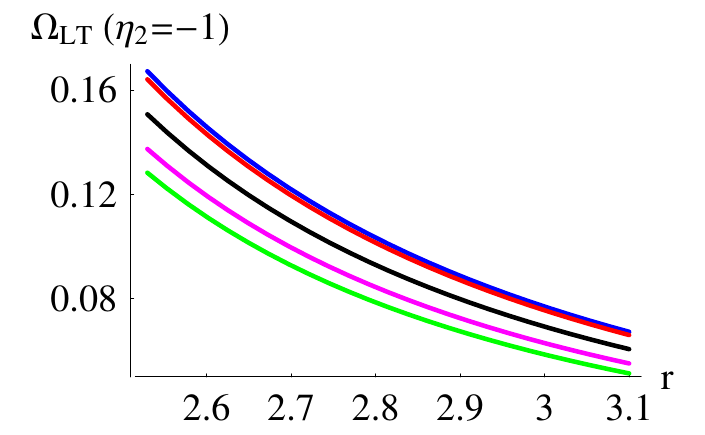}
	\caption{\footnotesize{Magnitude of the spin precession frequency $\Omega_{LT}$ versus $r$ for $M=1$, $\gamma=0$, $a=0.7$, $\theta=\pi/4$, and different values of $q$ and $\eta_2$. The blue plots correspond to $q =0.1$, the red plots to $q =0.3$, the black plots to $q =0.7$, the magenta plots to $q =1$, and the green plots to $q =1.2$.}} \label{FigO2}
\end{figure}

\begin{figure}[h!]
	\includegraphics[width=0.49\textwidth]{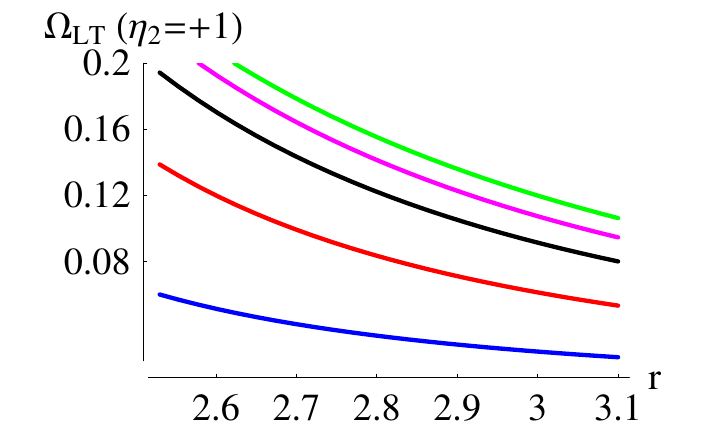}
	\includegraphics[width=0.49\textwidth]{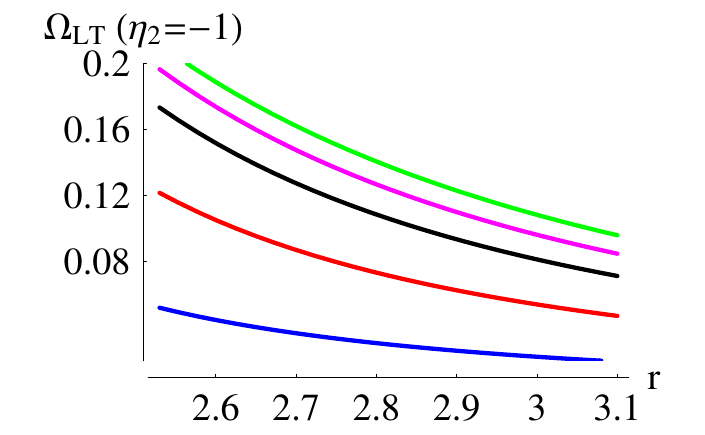}
	\caption{\footnotesize{Magnitude of the spin precession frequency $\Omega_{LT}$ versus $r$ for $M=1$, $\gamma=0$, $q=1/2$, $\theta=\pi/4$, and different values of $a$ and $\eta_2$. The blue plots correspond to $a =0.2$, the red plots to $a =0.5$, the black plots to $a =0.8$, the magenta plots to $a =1$, and the green plots to $a =1.2$.}} \label{FigO3}
\end{figure}

It is also obvious from Figs.~\ref{FigO2} and~\ref{FigO3}, where $\gamma=0$, that the Lense--Thirring effect is more enhanced for $\eta_2=+1$ if one fixes all the parameters but the charge as in Fig.~\ref{FigO2} or one fixes all the parameters but the rotation parameter as in Fig.~\ref{FigO3}. Again this concerns only the case $\gamma=0$; for $\gamma=1$ the effect is rather enhanced for $\eta_2=-1$ if one fixes the other parameters except the charge (not shown) or one fixes the other parameters except the rotation parameter (not shown).

A common feature is that the effect is enhanced with rotation whatever the value of the parameters are. In fact, the Lense--Thirring effects are actually due to rotation of central object, this is also verified for RPBH, as seen in Fig.~\ref{FigO3}, as the rotational parameter $a$ increases, the magnitude of precession frequency also increases. Further, the precession also depends on the direction along which the gyroscope is attached to the moving observer. At the equator (i.e. $\theta=\pi/2$), the frame dragging effect is maximum and it declines as the observer moves towards the pole. We have checked that $\Omega_{LT}$ increases with $\theta$ as the latter runs from 0 to $\pi/2$ and it assumes its maximum value for $\theta=\pi/2$ (not shown).

\section{Discussion\label{secdis}}
We have formulated and discussed the properties of rotating normal and phantom black holes of EMD theory, which, upon taking appropriate limits, reduce to the Kerr and Kerr--Newman solutions.

The presence of a phantom electromagnetic field ($\eta_2=-1$) and/or phantom scalar field  [$\eta_1=-1$ $\Leftrightarrow$ $\gamma\in (-\infty,-1)\cup[1,\infty)$] affects greatly the motion of photons and massive particles. Circular motion and the determination of isco and lsco are among the first tasks workers perform for purposes of astrophysical applications to accretion disks (see~\cite{Chandra,Teukolsky,Kerr1,Tursunov,Kerr2,Anew,Tursunov2,Tursunov3,accr1,accr2} and references terein). From this point of view we noticded that the presence of phantom fields may widen, in the $ra$-plane, the set of existence of stable circular orbits; in some other cases this same set is much restricted. We have shown that the effective potential and the Lense--Thirring precession frequency are greatly enhanced for some values of the phantom parameters.

The rotating solution derived in this work could have been obtained by mere substitution in the generic rotating solution derived in Refs.~\cite{A,A1}. Because of the importance of the procedure, which has known some applications~\cite{s1,s2,s3,s4,s5,s6,s7,s8,s9,s10,s11}, we have outlined it in a way slightly different from the presentation given in Refs.~\cite{A,A1}, in order to emphasize some arguments related with the determination of the still free three functions resulting from the transformation from Eddington--Finkelstein coordinates to Boyer--Lindquist coordinates. Further arguments are given in the last but one paragraph of Sec.~\ref{secrot}.

\end{document}